\documentclass{elsart}

\usepackage{graphicx}
\usepackage{amssymb}
\usepackage{amsmath}

\begin{document}

\begin{frontmatter}


%\title{}
%\author{}
%\ead{}
%\thanks[label1]{}
%\thanks[label2]{}

\title{Pairing and continuum effects on low-frequency quadrupole vibrations 
in deformed Mg isotopes close to the neutron drip line}

\author[label1]{K. Yoshida},
%\ead{kyoshida@ruby.scphys.kyoto-u.ac.jp}
\author[label2]{M. Yamagami}
\and
\author[label1]{K. Matsuyanagi}
\address[label1]{Department of Physics, Graduate School of Science, Kyoto University, 
Kyoto 606-8502, Japan}
\address[label2]{Radioactive Isotope Physics Laboratory, RIKEN, Wako Saitama 351-0198, Japan}

\begin{abstract}

Low-frequency quadrupole vibrational modes in deformed $^{36,38,40}$Mg 
close to the neutron drip line are studied by means of 
the quasiparticle-random-phase approximation 
based on the coordinate-space Hartree-Fock-Bogoliubov formalism. 
Strongly collective $K^{\pi}=0^{+}$ and $2^{+}$ excitation 
modes carrying $10-20$ Weisskopf units in the 
intrinsic isoscalar quadrupole transition strengths 
are obtained at about 3 MeV. 
There are two reasons for the enhancement of the transition strengths.
First, the quasiparticle wave functions generating these modes possess 
spatially very extended structure. 
The asymptotic selection rules characterizing the $\beta$ and $\gamma$ 
vibrations in stable deformed nuclei are thus strongly violated.
Second, the dynamic pairing effects act strongly to enhance the collectivity 
of these modes. It is suggested that the lowest $K^{\pi}=0^{+}$ collective mode 
is a particularly sensitive indicator of the nature of pairing 
correlations in deformed nuclei close to the neutron drip line.

\end{abstract}

\begin{keyword}
% keywords here, in the form: keyword \sep keyword \sep 
Hartree-Fock-Bogoliubov method \sep 
Quasiparticle-RPA \sep Collective excitations \sep  
Deformed unstable nuclei \sep 
Neutron drip line \sep Mg isotopes
\PACS 
21.60.Ev \sep 21.60.Jz \sep 21.10.Re
\end{keyword}
\end{frontmatter}

% main text

\section{Introduction}

The physics of drip-line nuclei is one of the current frontiers in nuclear structure physics~\cite{tan01,hor01,hag02a}. 
The number of unstable nuclei experimentally accessible will 
remarkably increase 
when the next generation of radioactive ion beam facilities start running. 
We shall be able to study the properties not only of the ground states
but also of low-lying excited states of drip-line nuclei in the medium-mass 
region. 
Collective excitation in neutron-rich nuclei is one of the most interesting 
issues in this field. 
Because properties of low-frequency collective vibrational modes are 
quite sensitive to
surface effects and details of shell structure, 
we expect that new kinds
of collective excitations emerge under such new situations of nuclear structure.
In order to quest for collective modes of excitation unique to 
unstable nuclei associated with new features such as neutron skins, 
many attempts have been made using the self-consistent RPA based on the 
Skyrme-Hartree-Fock (SHF) method~\cite{ham96,ham99,shl03} 
and the Quasiparticle-RPA (QRPA) 
including the pairing correlations~\cite{mat01,hag01,ben02,kha02,yam04,ter05}. 
A number of similar approaches using different mean fields have also been 
carried out~\cite{vre01,paa03,paa04,paa05,cao05,vre05,gia03,per05,sar04}. 
(See Refs.~\cite{ter05,vre05,ben03} for extensive lists of references 
concerning the self-consistent RPA and mean-field calculations.) 
Most of these calculations, however, are restricted to spherical nuclei. 

Quite recently,
low-frequency RPA modes in deformed nuclei close to the neutron drip line 
have been investigated by several authors.
The time-dependent Hartree-Fock method 
formulated in the three-dimensional coordinate space 
with a complex absorbing boundary condition 
was applied to low-frequency isovector dipole modes~\cite{nak05}. 
Possible appearance of low-frequency octupole vibrations  
built on superdeformed states in neutron drip-line nuclei was discussed 
in Ref.~\cite{ina05} 
on the basis of the SHF plus mixed representation RPA~\cite{lem68,mut02,ima03} 
calculations.
In Ref.~\cite{yos05}, we investigated properties of octupole 
excitations built on superdeformed states in neutron-rich sulfur isotopes 
by means of the RPA based on the deformed Woods-Saxon (WS) potential 
in the coordinate-space mesh-representation. 
We found that low-lying states created by excitation of a single neutron 
from a loosely bound low-$\Omega$ state to a high-$\Omega$ resonance state
($\Omega$ being the $z$-component of the angular momentum)  
acquire extremely strong octupole transition strengths 
due to very extended spatial structure of particle-hole wave functions.
All of these calculations, however, did not take into account the pairing 
correlation. 
In Refs.~\cite{urk01,alv04}, 
low-lying Gamow-Teller $\beta$-decay strengths were 
investigated by means of the proton-neutron RPA 
using the SHF + BCS approximation. 
Gamma vibrations in $^{38}$Mg were studied using the QRPA 
with the BCS approximation~\cite{hag04a} 
on the basis of the response function formalism. 
It should be noted that 
these calculations relied on the BCS approximation, which is inappropriate,  
because of the unphysical nucleon gas problem~\cite{dob84}, 
for describing continuum coupling effects in drip line nuclei. 

The nature of pairing correlations in neutron drip-line nuclei is 
one of the most important subjects in the physics of unstable nuclei. 
One of the unique features of drip-line nuclei is that 
the pairing correlation takes place not only among bound levels 
but also including continuum states. 
To describe this unique character of pairing, 
the coordinate-space Hartree-Fock-Bogoliubov (HFB) 
formalism is suitable~\cite{bul80,dob84} and has been widely used 
for the study of single-particle motion and shell structure 
near the continuum~\cite{smo93,dob94,dob96,ben99}. 
Due to the pairing and continuum effects, 
spatial structure of quasiparticle wave functions 
near the chemical potential changes significantly, 
which affects the properties of low-frequency excitation modes~\cite{yam05}. 
In order to study the effects of pairing on the low-frequency excitation modes 
in deformed nuclei near the neutron drip-line, 
we have extended the previous work to self-consistently include 
pairing correlations, and constructed 
a new computer code that carries out the deformed QRPA calculation  
on the basis of the coordinate-space HFB formalism. 

The aim of this paper is to carry out the deformed QRPA calculation 
for neutron drip-line nuclei and investigate 
the low-frequency quadrupole vibrational modes 
with $K^{\pi}=0^{+}$ and $2^{+}$
in ${}^{36,38,40}$Mg close to the neutron drip line.
According to the Skyrme-HFB calculations~\cite{ter97,sto03} 
and Gogny-HFB calculation~\cite{rod02}, 
these isotopes are well deformed. 
The shell-model calculation~\cite{cau04} also suggests that the ground state 
of ${}^{40}$Mg is dominated by the neutron two-particle-two-hole components, 
which is consistent with the breaking of the $N=28$ shell closure
discussed in~\cite{rei99}.
We investigate properties of low-frequency modes of excitation 
in these Mg isotopes simultaneously taking into account 
the deformed mean-field effects, 
the pairing correlations, and excitations into the continuum.

This paper is organized as follows: In the next section, 
the framework of the mean-field and QRPA calculations is briefly 
described. 
In Section 3, results of the RPA calculation for 
low-frequency quadrupole vibrations with $K^{\pi}=0^{+}$ and $2^{+}$ 
in $^{36,38,40}$Mg are presented and discussed focusing our attention to 
the microscopic mechanism of emergence of collective modes
in deformed superfluid nuclei close to the neutron drip line. 
Concluding remarks are given in \S4.

A preliminary version of this work was previously 
reported in Ref.~\cite{yos05_nil}.

\section{Method}

\subsection{Mean-field calculation}

In order to discuss simultaneously effects of nuclear deformation 
and pairing correlations including the continuum, 
we solve the HFB equation~\cite{bul80,dob84,obe03}
\begin{equation}
\begin{pmatrix}
h^{\tau}(\boldsymbol{r}\sigma)-\lambda^{\tau} & \tilde{h}^{\tau}(\boldsymbol{r}\sigma) \\
\tilde{h}^{\tau}(\boldsymbol{r}\sigma) & -(h^{\tau}(\boldsymbol{r}\sigma)-\lambda^{\tau}) \end{pmatrix}
\begin{pmatrix}
\varphi^{\tau}_{1,\alpha}(\boldsymbol{r}\sigma) \\ 
\varphi^{\tau}_{2,\alpha}(\boldsymbol{r}\sigma)
\end{pmatrix}
= E_{\alpha}
\begin{pmatrix}
\varphi^{\tau}_{1,\alpha}(\boldsymbol{r}\sigma) \\ 
\varphi^{\tau}_{2,\alpha}(\boldsymbol{r}\sigma)
\end{pmatrix} \label{eq:HFB1}
\end{equation}
directly in the cylindrical coordinate space 
assuming axial and reflection symmetry. 
In comparison to the conventional method of 
using a deformed harmonic oscillator basis, 
this method is more effective in the treatment of 
spatially extended wave functions, 
like loosely bound states, resonant states and continuum states. 
As is well known, 
when the quasiparticle energy $E$ is greater than the absolute magnitude 
$\vert\lambda\vert$ of the chemical potential, 
the upper component $\varphi_{1}(\boldsymbol{r}\sigma)$ 
obeys the scattering-wave boundary condition,
while the lower component  $\varphi_{2}(\boldsymbol{r}\sigma)$ 
is always exponentially decaying at infinity.
 
For the mean-field Hamiltonian $h$, we employ the deformed Woods-Saxon 
potential with the parameters used in~\cite{yos05}, 
except the isovector potential strength for which 
a slightly smaller value, 30 MeV in stead of 33 MeV, is adopted 
in order to describe $^{40}$Mg as a drip-line nucleus 
in accordance with the Skyrme-HFB~\cite{ter97,sto03} 
and Gogny-HFB calculations~\cite{rod02}.
The pairing field is treated self-consistently by using 
the density-dependent contact interaction~\cite{ber91,ter95}, 
\begin{equation}
v_{pp}(\boldsymbol{r},\boldsymbol{r}^{\prime})=V_{0}\dfrac{1-P_{\sigma}}{2}
\left[ 1- \eta \left( \dfrac{\varrho^{\mathrm{IS}}(\boldsymbol{r})}{\varrho_{0}}\right) \right]
\delta(\boldsymbol{r}-\boldsymbol{r}^{\prime}), \label{eq:res_pp}
\end{equation}
with $V_{0}=-450$ MeV$\cdot$fm$^{3}$ and $\varrho_{0}=0.16$ fm$^{-3}$,
where $\varrho^{\mathrm{IS}}(\boldsymbol{r})$ denotes the isoscalar density and 
$P_{\sigma}$ is the spin exchange operator. 
For the parameter $\eta$, which represents density dependence,
we use $\eta=1.0$(surface type). Sensitivity of calculated results 
to the parameter $\eta$ will be examined in subsection 3.4. 
The pairing Hamiltonian is then given by 
\begin{equation}
\tilde{h}^{\tau}(\boldsymbol{r})=\dfrac{V_{0}}{2}\left[ 
1-\eta \left( \dfrac{\varrho^{\mathrm{IS}}(\boldsymbol{r})}{\varrho_{0}}\right) \right]
\tilde{\varrho}^{\tau}(\boldsymbol{r}).
\label{pair_hamiltonian}
\end{equation}
The normal and abnormal (pairing) densities are given by
\begin{align}
\varrho^{\tau}(\rho,z)&=\sum_{\alpha}\sum_{\sigma=\pm1/2} 
|\varphi_{2,\alpha}^{\tau}(\rho,z,\sigma)|^{2}, \label{density} \\
\tilde{\varrho}^{\tau}(\rho,z)&=-\sum_{\alpha}\sum_{\sigma=\pm1/2}
\varphi_{1,\alpha}^{\tau}(\rho,z,\sigma)\varphi_{2,\alpha}^{\tau}(\rho,z,\sigma)
\label{pair_density}
\end{align}
and the mean-square radii of protons and neutrons are calculated as 
\begin{equation}
\langle r^{2} \rangle_{\tau} = 
\dfrac{\int \rho \mathrm{d}\rho \mathrm{d}z r^{2}\varrho^{\tau}(\rho,z)}
{\int \rho \mathrm{d}\rho \mathrm{d}z \varrho^{\tau}(\rho,z)}, 
\label{eq:neutron_radius}
\end{equation}
where $\boldsymbol{r}=(\rho,z)$, 
$r=\sqrt{\rho^{2}+z^{2}}$ and $\tau$=$\pi$ or $\nu$;
$\varrho^{\pi}(\rho,z)$ and $\varrho^{\nu}(\rho,z)$
being the proton and neutron densities. 
The average gaps are defined by~\cite{dob84,dob96} 
\begin{equation}
\langle \Delta_{\tau} \rangle=-\int \mathrm{d}\boldsymbol{r}
\varrho^{\tau}(\boldsymbol{r}) \tilde{h}^{\tau}(\boldsymbol{r})/\int \mathrm{d}\boldsymbol{r}
\varrho^{\tau}(\boldsymbol{r}).
\end{equation}

We construct the discretized Hamiltonian matrix 
by use of the finite difference method for derivatives 
and then diagonalize the matrix to obtain 
the quasiparticle wave functions on the two-dimensional lattice 
consisting of the cylindrical coordinates $\rho$ and $z$. 
The kinetic energy term and the spin-orbit potential 
are evaluated using the 9-point formula. 
Because the time-reversal symmetry and the reflection symmetry 
with respect to the $x-y$ plane are assumed, 
we have only to solve for positive $\Omega$ and positive $z$. 
We use the lattice mesh size $\Delta\rho=\Delta z=0.8$ fm 
and the box boundary condition at  
$\rho_{\mathrm{max}}=10.0$ fm and $z_{\mathrm{max}}=12.8$ fm. 
The quasiparticle energy is cut off at 50 MeV and 
the quasiparticle states up to $\Omega^{\pi}=13/2^{\pm}$ are included. 
This model space is larger than that used in Ref.~\cite{yos05_nil}. 
We impose the condition on the convergence of the pairing energy as 
$|(E^{(i)}_{pair}-E^{(i-1)}_{pair})/E^{(i)}_{pair}|<10^{-5}$,
where $i$ denotes the iteration number and 
the pairing energy is defined by \cite{dob96} 
\begin{equation}
E_{pair}=\dfrac{V_{0}}{4}\sum_{\tau=\pi,\nu} \int\mathrm{d}\boldsymbol{r}
\{\tilde{\varrho}^{\tau}(\boldsymbol{r})\}^{2}
\left[ 1-\eta \left(\dfrac{\varrho^{\mathrm{IS}}(\boldsymbol{r})}{\varrho_{0}}\right) \right].  
\end{equation}
We use the same deformation parameter $\beta_{2}=0.28$ 
in the Woods-Saxon potential for both neutrons and protons. 
This parameter is chosen to approximately reproduce the $Q-$moments 
calculated in Ref.\cite{ter97}. 
We checked that properties of the QRPA modes do not change significantly 
when the deformation parameter is varied around $\beta_{2}\sim 0.3$.

\subsection{Quasiparticle-RPA calculation}

Using the quasiparticle basis obtained in the previous subsection,  
we solve the QRPA equation in the standard matrix formulation~\cite{row70} 
\begin{equation}
\sum_{\gamma \delta}
\begin{pmatrix}
A_{\alpha \beta \gamma \delta} & B_{\alpha \beta \gamma \delta} \\
B_{\alpha \beta \gamma \delta} & A_{\alpha \beta \gamma \delta}
\end{pmatrix}
\begin{pmatrix}
f_{\gamma \delta}^{\lambda} \\ g_{\gamma \delta}^{\lambda}
\end{pmatrix}
=\hbar \omega_{\lambda}
\begin{pmatrix}
1 & 0 \\ 0 & -1
\end{pmatrix}
\begin{pmatrix}
f_{\alpha \beta}^{\lambda} \\ g_{\alpha \beta}^{\lambda}
\end{pmatrix} \label{eq:AB1}.
\end{equation}
This method is convenient to analyze microscopic structures of 
the QRPA eigenmodes in comparison with other RPA formalisms 
based on the Greens function method. 

The residual interactions in the particle-particle channel 
appearing in the QRPA matrices $A$ and $B$ are self-consistently treated 
using the density-dependent contact interaction (\ref{eq:res_pp}).
On the other hand, 
for residual interactions in the particle-hole channel, 
we use the Skyrme-type interaction~\cite{sho75} 
\begin{equation}
v_{ph}(\boldsymbol{r},\boldsymbol{r}^{\prime})=
\left[ t_{0}(1+x_{0}P_{\sigma})+\dfrac{t_{3}}{6}(1+x_{3}P_{\sigma})
\varrho^{\mathrm{IS}}(\boldsymbol{r}) \right]
\delta(\boldsymbol{r}-\boldsymbol{r}^{\prime}), \label{eq:res_ph}
\end{equation}
with $t_{0}=-1100$ MeV$\cdot$fm$^{3}$, $t_{3}=16000$ MeV$\cdot$fm$^{6}, 
x_{0}=0.5$, and $x_{3}=1.0$.
Because the deformed Wood-Saxon potential is used for the mean-field, 
we renormalize the residual interaction in the particle-hole channel 
by multiplying a factor $f_{ph}$
to get the spurious $K^{\pi}=1^{+}$  mode
(representing the rotational mode) at zero energy
($v_{ph} \rightarrow f_{ph}\cdot v_{ph}$). 
We cut the model space for the QRPA calculation by
$E_{\alpha}+E_{\beta} \leq 30$MeV,  which is 
smaller than that for the HFB calculation. 
Accordingly, we need another self-consistency factor $f_{pp}$
for the particle-particle channel. 
We determine this factor such that the spurious $K^{\pi}=0^{+}$ mode 
associated with the number fluctuation appears at zero energy
($v_{pp} \rightarrow f_{pp}\cdot v_{pp}$). 
In the present calculation, 
the dimension of QRPA matrix is about 3700 
for the $K^{\pi}=0^{+}$ modes in $^{40}$Mg. 

In terms of the nucleon annihilation and creation operators 
in the coordinate representation, $\hat{\psi}(\boldsymbol{r}\sigma)$ 
and $\hat{\psi}^{\dagger}(\boldsymbol{r}\sigma)$, 
the quadrupole operator is represented as 
$\hat{Q}_{2K}
=\sum_{\sigma}\int \mathrm{d}\boldsymbol{r}r^{2}Y_{2K}(\hat{r})
\hat{\psi}^{\dagger}(\boldsymbol{r}\sigma)
\hat{\psi}(\boldsymbol{r}\sigma)$.
The intrinsic matrix elements 
$\langle \lambda|\hat{Q}_{2K}|0 \rangle$ 
of the quadrupole operator between the excited state $|\lambda \rangle$ 
and the ground state $|0\rangle$ are given by
\begin{equation}
\langle \lambda|\hat{Q}_{2K}|0 \rangle=\sum_{\alpha \beta}
Q_{2K,\alpha \beta}^{(\mathrm{uv})} 
(f_{\alpha \beta}^{\lambda}+g_{\alpha \beta}^{\lambda})
=\sum_{\alpha \beta}M_{2K,\alpha \beta}^{(\mathrm{uv})} 
\label{eq:matrix_element},
\end{equation}
where 
\begin{equation}
Q_{2K,\alpha \beta}^{(\mathrm{uv})}
\equiv 2\pi\delta_{K,\Omega_{\alpha}+\Omega_{\beta}}
\int\!\!\mathrm{d}\rho\mathrm{d}z Q_{2K,\alpha \beta}^{(\mathrm{uv})}(\rho,z),
\label{eq:integrand} 
\end{equation} 
with
\begin{multline}
Q_{2K,\alpha \beta}^{(\mathrm{uv})}(\rho,z)
=\rho \{ \varphi_{1,\alpha}(\rho,z,\downarrow)\varphi_{2,\beta}(\rho,z,\uparrow)
-\varphi_{1,\alpha}(\rho,z,\uparrow)\varphi_{2,\beta}(\rho,z,\downarrow) \\
-\varphi_{1,\beta}(\rho,z,\downarrow)\varphi_{2,\alpha}(\rho,z,\uparrow) 
+\varphi_{1,\beta}(\rho,z,\uparrow)\varphi_{2,\alpha}(\rho,z,\downarrow) \} Q_{2K}(\rho,z).
\end{multline}
Here $Q_{2K}(\rho,z)=Q_{2K}(\boldsymbol{r})e^{-iK\varphi}=r^{2}Y_{2K}(\theta,\varphi)e^{-iK\varphi}$.

We calculate the transition strength functions 
\begin{equation}
S^{\mathrm{IS}}(\omega)
=\sum_{\lambda}|\langle \lambda|\hat{Q}^{\mathrm{IS}}_{2K}|0 \rangle|^{2} 
\delta(\hbar\omega-\hbar\omega_{\lambda})
\end{equation}
for isoscalar quadrupole operators $\hat{Q}^{\mathrm{IS}}_{2K}
=\hat{Q}^{\pi}_{2K}+\hat{Q}^{\nu}_{2K}$, 
and use notations $B(Q^{\tau}2)=
|\langle \lambda|\hat{Q}^{\tau}_{2K}|0 \rangle|^{2}$ 
for transition strengths 
and $M_{\tau}=\langle \lambda | \hat{Q}^{\tau}_{2K}|0 \rangle$ 
for transition matrix elements ($\tau=\pi, \nu$, IS). 
Note that these quantities are defined in the intrinsic coordinate frame
associated with the deformed mean field, so that 
appropriate Clebsh-Gordan coefficients should be 
multiplied to obtain transition probabilities 
in the laboratory frame~\cite{BM2}.
For instance, a factor 1/5 should be multiplied 
for obtaining the transition strength
$B(E2; 2^{+}_{1} \to 0^{+}_{\beta})$
from the $2_{1}^{+}$ state to the $0^{+}_{\beta}$ state, while 
the factor is unity for obtaining the transition strength 
$B(E2; 0^{+}_{\mathrm{gs}} \to 2^{+}_{\beta})$
from the ground state to the $2^{+}_{\beta}$ state 
built on the excited $K^{\pi}=0^{+}$ state. 
Here, $2_{1}^{+}$ denotes the $2^{+}$ member of the ground-state rotational
band, while $0^{+}_{\beta}$ and $2^{+}_{\beta}$ indicate the 
rotational band members associated with the $K^{\pi}=0^{+}$ intrinsic 
excitations. 

\section{Results and Discussion}
\subsection{Some features of calculated results}
\begin{figure}[tp]
\begin{center}
\includegraphics[scale=1.05]{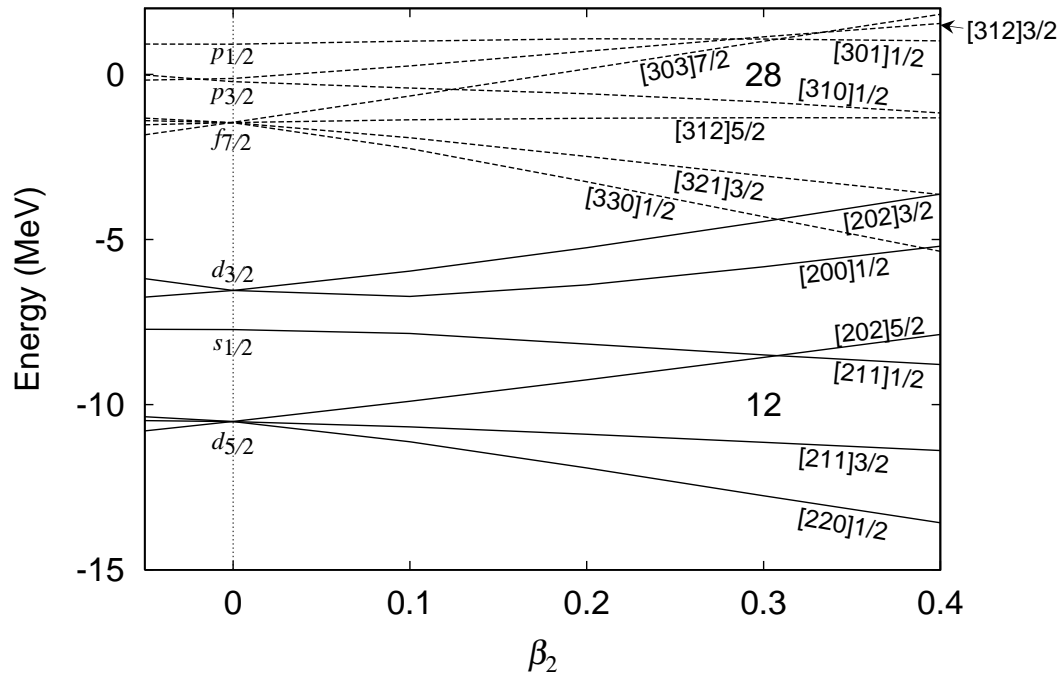}
\caption{Single-particle energies in the deformed WS potential 
for neutrons in $^{40}$Mg, 
plotted as functions of the quadrupole deformation parameter $\beta_{2}$. 
Solid and dotted lines denote positive- and 
negative-parity levels, respectively. 
Single-particle levels are labeled with 
the asymptotic quantum numbers $[Nn_{3}\Lambda]\Omega$.
}
\label{40Mg_nilsson}
\end{center}
\end{figure}

The single-particle shell structure around the Fermi surface for neutrons 
in $^{36,38,40}$Mg exhibits an interesting feature.
Figure~\ref{40Mg_nilsson} shows the single-particle energy diagram 
for the WS potential as functions of deformation parameter $\beta_{2}$. 
As $\beta_{2}$ increases, a level crossing between the up-sloping [303]7/2 
level and the down-sloping [310]1/2 level takes place, and a deformed shell gap
is formed at $N=28$ around $\beta_{2}=0.3$. This deformed closed shell 
approximately corresponds to the $(f_{7/2})^{-2}(p_{3/2})^{2}$ configuration
in the spherical shell model representation.
The highest occupied level in this deformed closed shell is situated very near 
to the continuum threshold, so that there is no bound level above it.
However, neutron particle-hole excitations may take place 
into resonance levels like [303]7/2, [301]1/2 [312]3/2 
lying just above the continuum threshold.
In fact, as we shall discuss below, these resonance levels participate 
in the pairing correlations and play an important role in generating 
low-frequency collective modes of excitation in $^{36,38,40}$Mg. 
Thus, $^{40}$Mg and its neighboring isotopes provide an interesting situation to 
investigate collective modes unique in unstable nuclei 
near the neutron drip line.  
The resonance character of these levels just above the continuum threshold 
is confirmed by means of the eigenphase-sum method (see Appendix).

\begin{table}[tp]
\caption{Ground state properties of $^{36,38,40}$Mg obtained by the
deformed WS-HFB calculation with $\beta_2=0.28$. 
Chemical potentials, average pairing gaps, and 
root-mean-square radii for protons and neutrons are listed.
}
\label{GroundState}
\begin{center} 
\begin{tabular}{c||c|c|c||c|c|c}
 & $\lambda_{\pi}$ & $\overline{\Delta}_{\pi}$ & $\sqrt{\langle r^{2} \rangle_{\pi}}$ & 
$\lambda_{\nu}$ & $\overline{\Delta}_{\nu}$ & $\sqrt{\langle r^{2} \rangle_{\nu}}$  \\
nucleus  & (MeV) &  (MeV) &  (fm) & (MeV) &  (MeV) &  (fm) 
\\ \hline \hline
$^{36}$Mg & $-20.0$ & 0.0 & 3.06 & $-2.09$ & 1.81 & 3.74 \\
$^{38}$Mg & $-23.0$ & 0.0 & 3.08 & $-1.15$ & 1.98 & 3.86 \\
$^{40}$Mg & $-25.1$ & 0.0 & 3.10 & $-0.41$ & 2.14 & 3.99 \\
\end{tabular}
\end{center} 
\end{table}

Results of the deformed WS plus HFB calculation for the ground state 
properties of $^{36,38,40}$Mg are listed in Table~\ref{GroundState}.
Calculated values of the average pairing gap for neutrons are rather 
close to the value estimated in terms of the conventional systematics 
$\Delta_{\rm syst}=12/\sqrt{A}\simeq 1.9$MeV.
On the other hand, the average pairing gaps for protons vanish. 
As shown in this table, 
the neutron root-mean-square radius increases 
as approaching the neutron drip line, 
while the proton root-mean-square radius remains almost constant. 
This means that the neutron skin structure 
emerges in these nuclei; the difference between the neutron and proton radii 
in $^{40}$Mg is about 0.9 fm. 
 
\begin{figure}[bp]
\begin{center}
\includegraphics[scale=0.98]{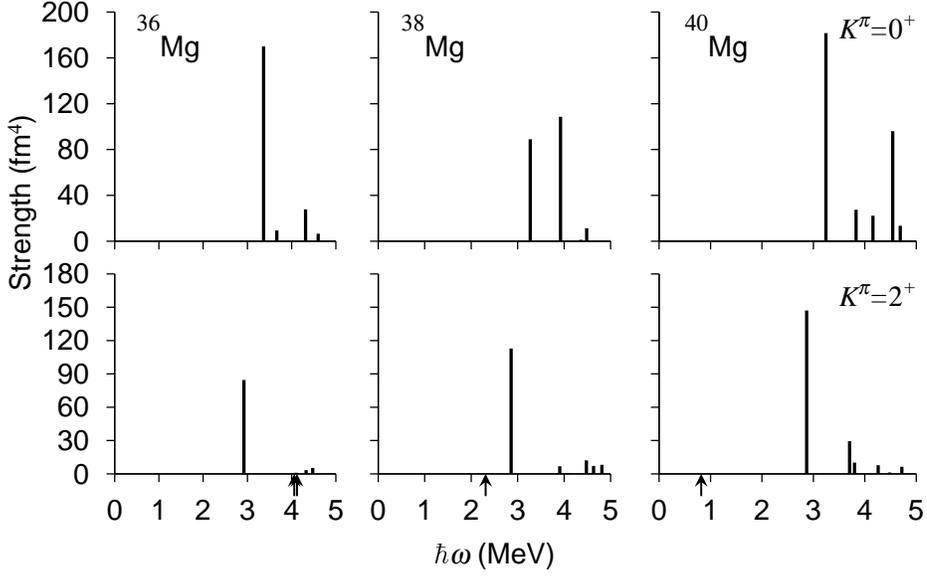}
\caption{Isoscalar quadrupole transition strengths $B(Q^{\mathrm{IS}}$2) 
for the $K=0^{+}$ excitations (upper panel) and the $K=2^{+}$ excitations 
(lower panel) 
built on the prolately deformed ground states of ${}^{36,38,40}$Mg.
The arrows beside the abscissa axes indicate the neutron threshold energies, 
$E_{\mathrm{th}}=4.06$ MeV 
(one-quasiparticle (1qp) continuum; $|\lambda|+\min E_{\alpha}$), 
4.12 MeV (two quasiparticle (2qp) continuum; $2|\lambda|$) for ${}^{36}$Mg, 
2.31 MeV (2qp continuum) for ${}^{38}$Mg and 
0.82 MeV (2qp continuum) for ${}^{40}$Mg. 
The QRPA calculations are made by using the surface-type pairing 
interaction and $\beta_{2}=0.28$ for both protons and neutrons.
}
\label{Mg_strength}
\end{center}
\end{figure}

Results of the QRPA calculation for quadrupole transition strengths 
are displayed in Fig.~\ref{Mg_strength}. 
We see prominent peaks at about 3 MeV for both
the $K^{\pi}=0^{+}$ and $2^{+}$ excitations. 
Their strengths are much larger than the single-particle strengths 
indicating collective character of these excitations.
The strength of the lowest $K^{\pi}=2^{+}$ excitation
gradually increases as approaching the neutron drip line,
while the lowest $K^{\pi}=0^{+}$ excitations in $^{36}$Mg and $^{40}$Mg
seem to be split into two peaks in the case of $^{38}$Mg.
In the following, we make an extensive analysis on 
microscopic structure of these low-frequency collective excitations.

\subsection{$K^{\pi}=0^{+}$ modes}

\begin{figure}[tp]
  \begin{center}
    \begin{tabular}{cc}
	\includegraphics[height=8cm]{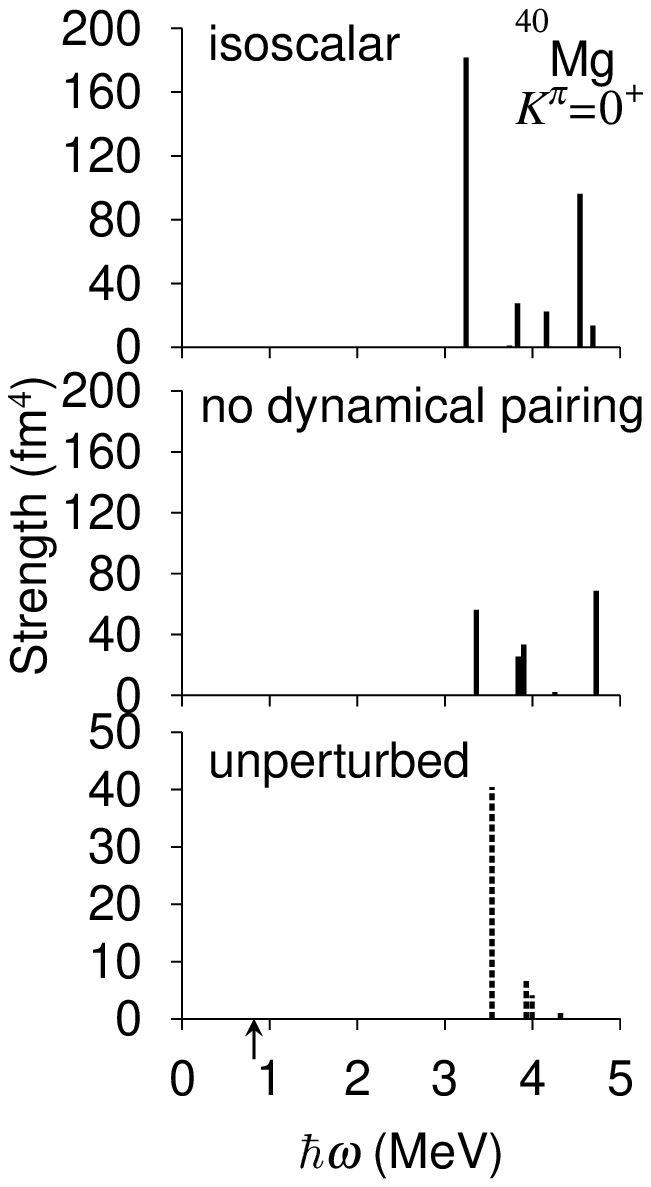}
	\hspace{0.7cm}
	\includegraphics[height=8cm]{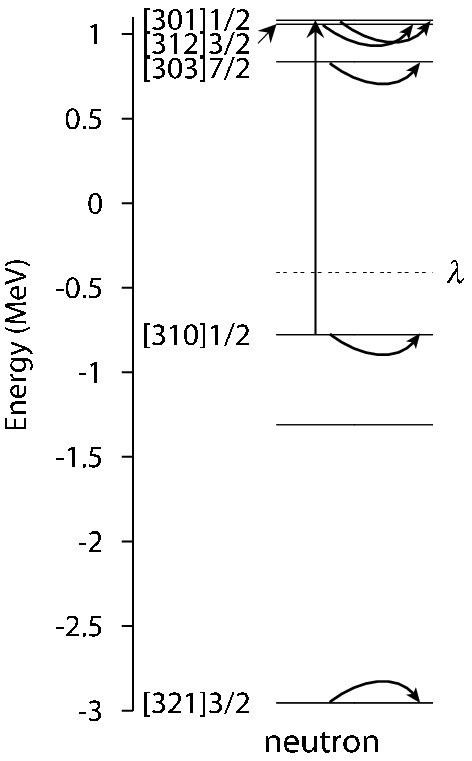}
    \end{tabular}
\caption{{\it Left}: 
Isoscalar quadrupole transition strengths $B(Q^{\mathrm{IS}}2$) 
for the $K^{\pi}=0^{+}$ excitations in ${}^{40}$Mg.
Results of the QRPA calculation with and without including 
the dynamical pairing effects are plotted in the upper and middle panels, 
respectively, while unperturbed two-quasiparticle strengths 
are shown in the lower panel. Notice that different scale is used for 
the unperturbed strengths.
The arrow beside the abscissa axis indicates 
the neutron threshold energy $2|\lambda|=0.82$ MeV. 
{\it Right}: 
Two-quasiparticle excitations generating 
the lowest $K^{\pi}=0^{+}$ mode at 3.2 MeV. 
The single-particle levels for the deformed WS potential are 
labeled with the asymptotic quantum numbers $[Nn_{3}\Lambda]\Omega$. 
The chemical potential $\lambda$ is indicated by the dashed line.
}
\label{40Mg_0+}
  \end{center}
\end{figure}

\begin{table}[tbp]
\caption{
QRPA amplitudes of the $K^{\pi}=0^{+}$ mode at 3.2 MeV in $^{40}$Mg.
This mode has $B(E2)=3.4 ~e^{2}$fm$^{4}$, 
$B(Q^{\nu}$2)=136 ~fm$^{4}$, and $B(Q^{\mathrm{IS}}2)=182$~fm$^{4}$.
The single-particle levels are labeled with
the asymptotic quantum numbers $[Nn_{3}\Lambda]\Omega$
of the dominant components of the wave functions. 
Only components with $|f_{\alpha \beta}|^{2}-|g_{\alpha\beta}|^{2} > 0.01$ are listed. 
}
\label{40Mg_amplitude0+}
\begin{center} 
\begin{tabular}{c|c|c|c|c|c|c}
 & $\alpha$ & $\beta$ & $E_{\alpha}+E_{\beta}$ & $|f_{\alpha \beta}|^{2}-|g_{\alpha\beta}|^{2}$ & 
$Q_{20,\alpha\beta}^{(\mathrm{uv})}$ & $M_{20,\alpha\beta}^{(\mathrm{uv})}$ \\
 &  &  & (MeV) &  & (fm$^{2}$) & (fm$^{2}$)  \\ \hline \hline
(a) & $\nu$[310]1/2 & $\nu$[310]1/2 & 3.54 & 0.438 & 6.36 & 4.27  \\
(b) & $\nu$[301]1/2 & $\nu$[310]1/2 & 3.93 & 0.067 & $-2.57$ & 0.925  \\ 
(c) & $\nu$[312]3/2 & $\nu$[312]3/2 & 3.99 & 0.280 & $-2.03$ & 1.08  \\
(d) & $\nu$[301]1/2 & $\nu$[301]1/2 & 4.32 & 0.027 & $0.992$ & $-0.176$  \\
(e) & $\nu$[303]7/2 & $\nu$[303]7/2 & 5.76 & 0.077 & $-3.39$ & 0.966  \\
(f) & $\nu$[321]3/2 & $\nu$[321]3/2 & 7.15 & 0.011 & 3.23 & 0.396  \\
\end{tabular}
\end{center} 
\end{table}

We first discuss the $K^{\pi}=0^{+}$ excitation modes in $^{40}$Mg. 
The QRPA transition strengths are compared with  unperturbed 
two-quasiparticle strengths in Fig.~\ref{40Mg_0+} .
A prominent peak is seen at about 3.2 MeV in the QRPA strength distribution; 
it possesses an enhanced strength of about 22 Weisskopf unit 
(1 W.u. $\simeq$ 8.1 fm$^{4}$ for $^{40}$Mg). 
From the QRPA amplitudes listed in Table~\ref{40Mg_amplitude0+}, 
it is clear that this collective mode is generated by coherent superposition 
of neutron excitations of both particle-hole and particle-particle types.
In Fig.~\ref{40Mg_0+}, the QRPA strengths are also compared 
with the strengths without the dynamical pairing effects, i.e., 
the result of QRPA calculation ignoring the residual pairing interactions. 
One immediately notice that the transition strength 
to the lowest excited state is drastically reduced when the 
dynamical pairing effects are ignored.

Let us discuss the reason why the lowest $K^{\pi}=0^{+}$ mode acquires 
eminently large transition strength. 
There are two points to understand this mechanism:
1) existence of unperturbed two-quasiparticle configurations 
possessing large transition strengths, and 
2) effect of residual interactions producing coherence 
among various two-quasiparticle configurations.

\begin{figure}[tp]
\begin{center}
\includegraphics[scale=0.85]{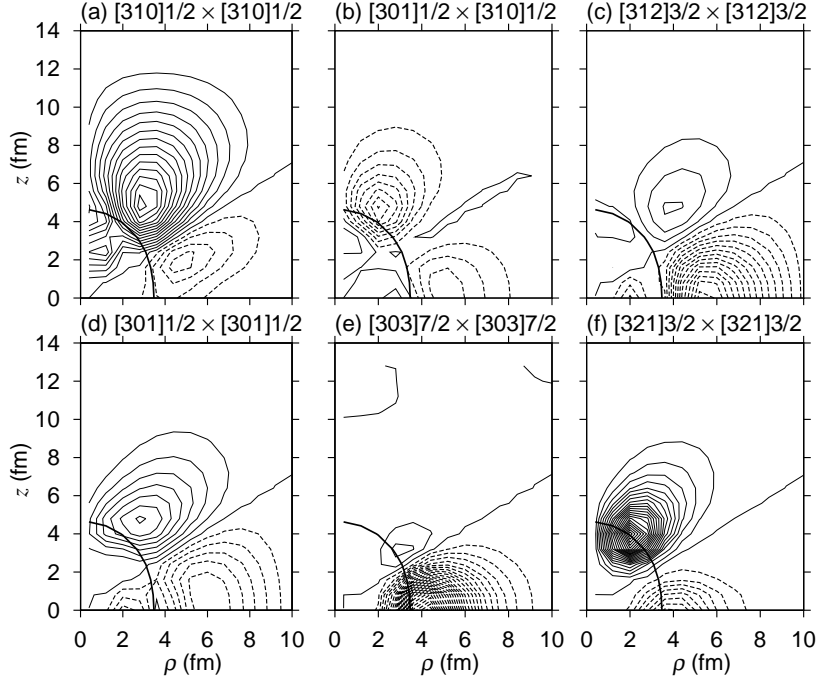}
\caption{
Spatial distribution functions $Q_{20,\alpha \beta}^{(\mathrm{uv})}(\rho,z)$ 
for some two-quasiparticle excitations 
generating the lowest $K^{\pi}=0^{+}$ mode in $^{40}$Mg.  
The contour lines are plotted at intervals of 0.002. 
The solid and dashed lines represent positive and negative quantities, 
respectively.
The thick solid line indicates the neutron half-density radius; 
$\varrho_{\nu}(0)/2 \sim 0.045$fm$^{-3}$.}
\label{integrand}
\end{center}
\end{figure}

To examine the first point, we plot in Fig.~\ref{integrand} 
spatial distributions of the quadrupole transition amplitudes for 
major two-quasiparticle configurations generating 
the lowest $K^{\pi}=0^{+}$ mode.  
We see that they are notably extended beyond the half-density radius.
This is a situation similar to that encountered in Ref.~\cite{yos05}, 
where a neutron excitation from a loosely bound state to 
a resonance state brings about very large transition strength.
We also note that the transition strength associated with the 
$\nu[301]1/2 \otimes \nu[310]1/2$ configuration is 
much enhanced although it should be hindered if 
the selection rule $\Delta N=2$ for the asymptotic quantum numbers 
is applied. This selection rule is broken for 
matrix elements associated with loosely bound states,
because their radial wave functions are spatially extended 
and quite different from those of the the harmonic oscillator potential. 

Concerning the second point, we have found that the dynamical pairing plays 
an especially important role. This point is easily seen by comparing 
the QRPA calculations with and without the dynamical pairing effects, 
which are shown in Fig.~\ref{40Mg_0+}. 
It is apparent that the prominent lowest peak disappears 
when the dynamical paring effects are ignored. 
We can say that the coherent superpositions among the particle-hole, 
particle-particle and hole-hole excitations
are indispensable for the emergence of this mode.
The importance of the coupling between 
the (particle-hole type) $\beta$ vibration and 
the (particle-particle and hole-hole type) pairing vibration 
has been well known in stable deformed nuclei~\cite{BM2}.
A new feature of the $K^{\pi}=0^{+}$ mode in neutron drip-line nuclei 
under discussion is that this coupling takes place among two-quasiparticle 
configurations that are loosely bound or resonances, 
so that their transition strengths are strikingly enhanced. 
In addition,  as seen in Fig.~\ref{integrand}, 
their spatial structures (peak positions and distribution) 
are rather similar with each other.
This is a favorable situation to generates coherence among them~\cite{yam05}.
The importance of dynamical pairing effects in generating soft dipole 
excitations has been demonstrated by Matsuo et al.~\cite{mat05} 
for spherical unstable nuclei near the neutron drip line.

\begin{figure}[tp]
\begin{center}
\begin{tabular}{cc}
\includegraphics[height=8cm]{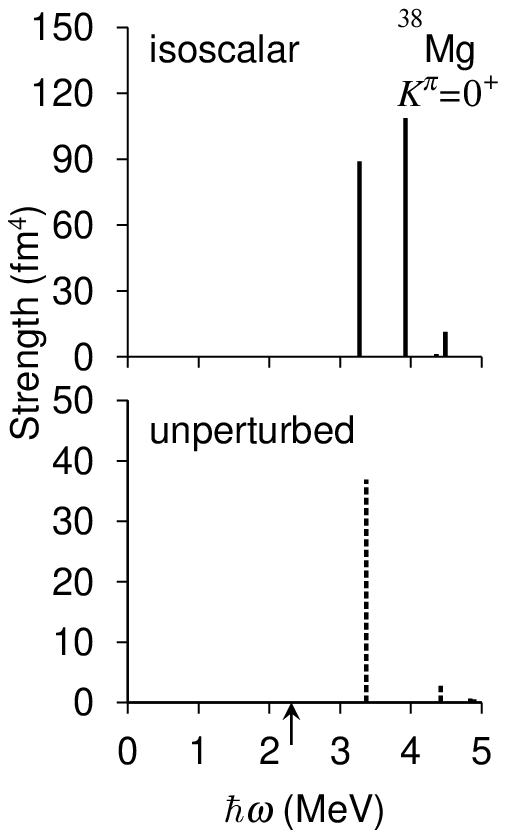}
\includegraphics[height=8cm]{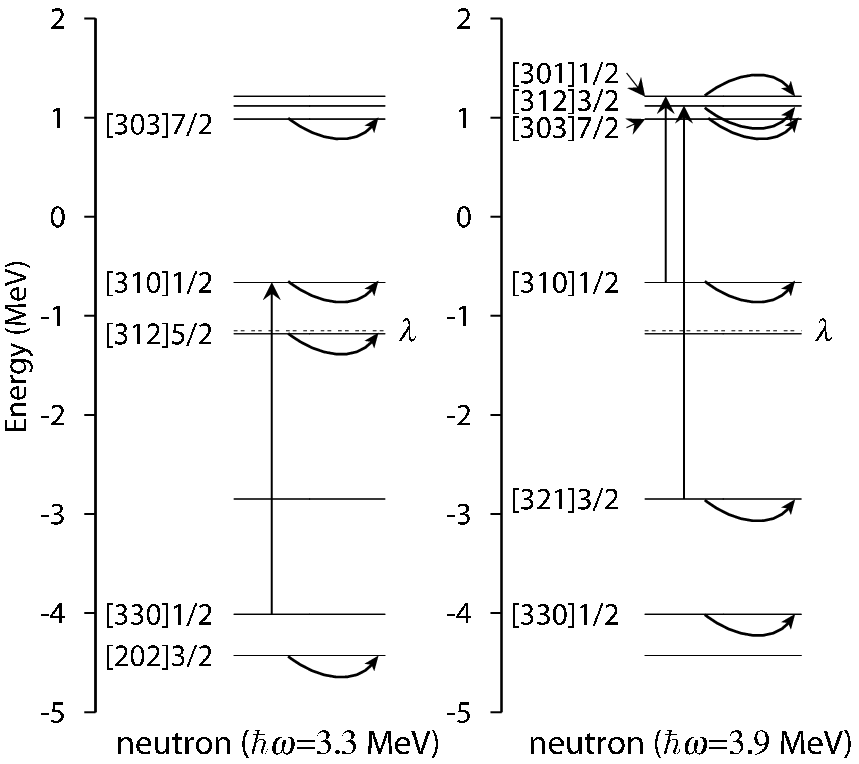} 
\end{tabular}
\end{center}
\caption{
{\it Left}: 
Isoscalar quadrupole transition strengths $B(Q^{\mathrm{IS}}2)$ 
for the $K^{\pi}=0^{+}$ excitations in ${}^{38}$Mg 
are plotted in the upper panel, while unperturbed two-quasiparticle strengths 
are shown in the lower panel.  
The arrow beside the abscissa axis indicates 
the neutron threshold energy $2|\lambda|=2.31$ MeV.
{\it Right}: 
Two-quasiparticle excitations generating 
the low-lying $K^{\pi}=0^{+}$ modes at 3.3 MeV and 3.9 MeV.
}
\label{38Mg_0+}
\end{figure}

\begin{table}[tp]
\caption{QRPA amplitudes of the $K^{\pi}=0^{+}$ mode at 3.3 MeV in $^{38}$Mg. 
This mode has $B(E2)=1.67 ~e^{2}$fm$^{4}$, 
$B(Q^{\nu}$2)=66.3 fm$^{4}$, $B(Q^{\mathrm{IS}}$2)=89.0 fm$^{4}$, 
and $\sum|g_{\alpha\beta}|^{2}=2.32 \times 10^{-2}$. 
Only components with $|f_{\alpha\beta}|^{2}-|g_{\alpha\beta}|^{2} > 0.01$ 
are listed. 
}
\label{38Mg_amplitude0+_1}
\begin{center} 
\begin{tabular}{c|c|c|c|c|c|c}
 & $\alpha$ & $\beta$ & $E_{\alpha}+E_{\beta}$ & $|f_{\alpha \beta}|^{2}-|g_{\alpha\beta}|^{2}$ & 
$Q_{20,\alpha\beta}^{(\mathrm{uv})}$ & $M_{20,\alpha\beta}^{(\mathrm{uv})}$ \\
 &  &  & (MeV) &  & (fm$^{2}$) & (fm$^{2}$) \\ \hline \hline
(a) & $\nu$[310]1/2 & $\nu$[310]1/2 & 3.37 & 0.673 & 6.08 & 5.25 \\
(b) & $\nu$[312]5/2 & $\nu$[312]5/2 & 4.84 & 0.146 & 0.821 & $-0.293$ \\
(c) & $\nu$[310]1/2 & $\nu$[330]1/2 & 5.35 & 0.023 & $-3.59$ & 0.769  \\
(d) & $\nu$[303]7/2 & $\nu$[303]7/2 & 6.35 & 0.066 & $-2.64$ & 0.614 \\
(e) & $\nu$[202]3/2 & $\nu$[202]3/2 & 7.82 & 0.021 & $-1.29$ & 0.149 \\
\end{tabular}
\end{center} 
\end{table}

\begin{table}[tp]
\caption{QRPA amplitudes of the $K^{\pi}=0^{+}$ mode at 3.9 MeV in $^{38}$Mg. 
This mode has $B(E2)=4.72 ~e^{2}$fm$^{4}$, 
$B(Q^{\nu}$2)=68.1 fm$^{4}$, $B(Q^{\mathrm{IS}}$2)=109 fm$^{4}$, 
and $\sum|g_{\alpha\beta}|^{2}=2.71 \times 10^{-2}$. 
Only components with $|f_{\alpha\beta}|^{2}-|g_{\alpha\beta}|^{2} > 0.01$ 
are listed. 
}
\label{38Mg_amplitude0+_2}
\begin{center} 
\begin{tabular}{c|c|c|c|c|c|c}
 & $\alpha$ & $\beta$ & $E_{\alpha}+E_{\beta}$ & $|f_{\alpha \beta}|^{2}-|g_{\alpha\beta}|^{2}$ & 
$Q_{20,\alpha\beta}^{(\mathrm{uv})}$ & $M_{20,\alpha\beta}^{(\mathrm{uv})}$ \\
 &  &  & (MeV) &  & (fm$^{2}$) & (fm$^{2}$) \\ \hline \hline
(a) & $\nu$[310]1/2 & $\nu$[310]1/2 & 3.37 & 0.037 & 6.08 & 1.34 \\
(b) & $\nu$[301]1/2 & $\nu$[310]1/2 & 4.42 & 0.258 & 1.67 & $-1.20$  \\
(c) & $\nu$[312]3/2 & $\nu$[312]3/2 & 4.90 & 0.048 & 0.716 & 0.169  \\
(d) & $\nu$[312]3/2 & $\nu$[321]3/2 & 5.47 & 0.250 & $-3.04$ & $-2.20$ \\
(e) & $\nu$[301]1/2 & $\nu$[301]1/2 & 5.47 & 0.018 & 0.802 & 0.131  \\
(f) & $\nu$[321]3/2 & $\nu$[321]3/2 & 6.04 & 0.058 & 1.66 & $-0.411$ \\
(g) & $\nu$[303]7/2 & $\nu$[303]7/2 & 6.35 & 0.084 & $-2.64$ & $-0.853$ \\
(h) & $\nu$[330]1/2 & $\nu$[330]1/2 & 7.33 & 0.099 & 4.57 & $-1.48$  \\ 
\end{tabular}
\end{center} 
\end{table}

Next we discuss the $K^{\pi}=0^{+}$ excitations in $^{38}$Mg and $^{36}$Mg.
The quadrupole transition strengths calculated for $^{38}$Mg 
are presented in Fig.~\ref{38Mg_0+},  
which exhibits two peaks below 4 MeV. 
The major two-quasiparticle excitations generating these peaks 
are illustrated in the middle and right panels of this figure. 
Their QRPA amplitudes  are listed in Tables~\ref{38Mg_amplitude0+_1} 
and \ref{38Mg_amplitude0+_2}. From these Tables, it is seen that  
the peak at 3.3 MeV is mainly generated
by the particle-particle type 
$\nu[310]1/2 \otimes \nu[310]1/2$ and  
$\nu[312]5/2 \otimes \nu[312]5/2$ excitations,  
while the peak at 3.9 MeV is mainly associated with 
the particle-hole type  
$\nu[301]1/2 \otimes \nu[310]1/2$ and 
$\nu[312]3/2 \otimes \nu[321]3/2$ excitations. 

\begin{figure}[bp]
  \begin{center}
    \begin{tabular}{cc}
	\includegraphics[height=8cm]{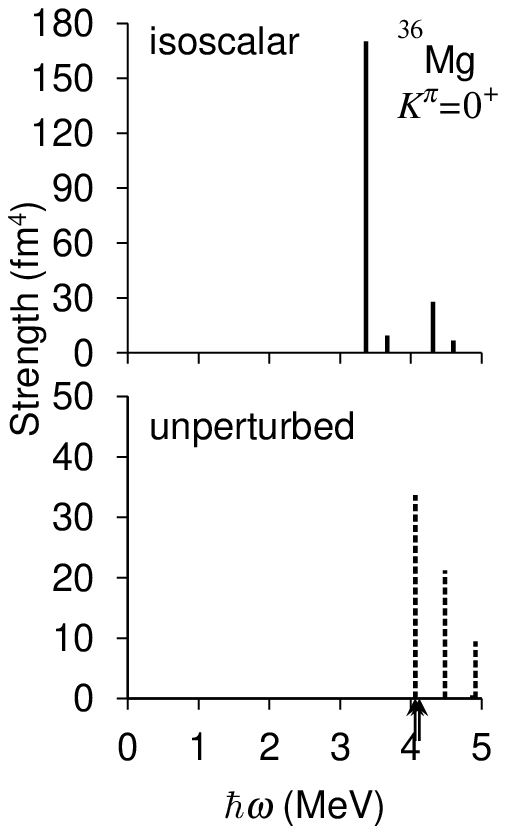}
	\includegraphics[height=8cm]{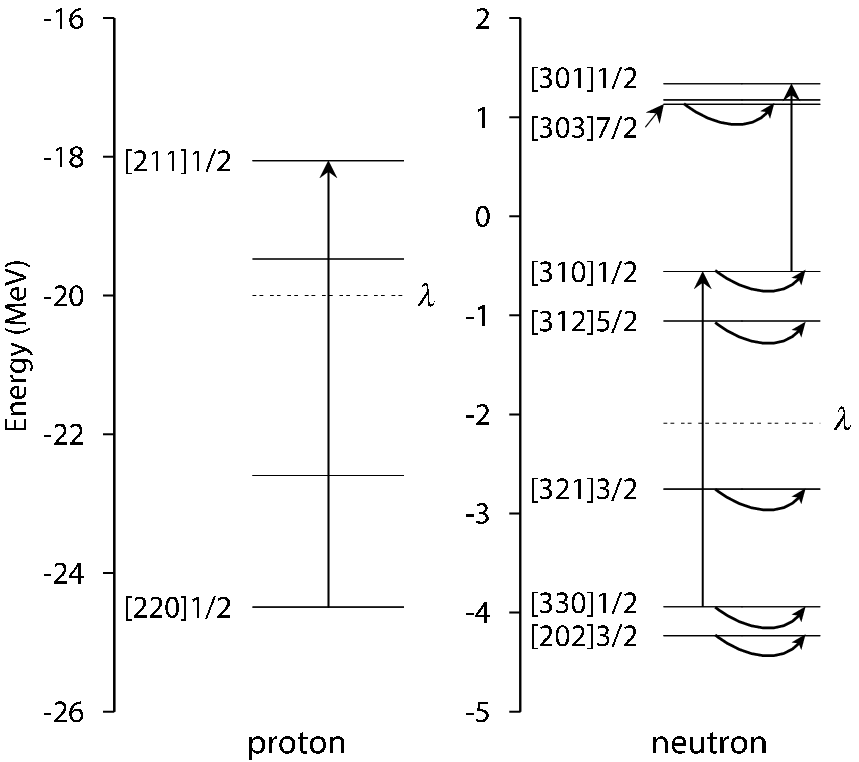}
    \end{tabular}
\caption{{\it Left}: 
Isoscalar quadrupole transition strengths 
$B(Q^{\mathrm{IS}}2)$ for the $K^{\pi}=0^{+}$ 
excitations in ${}^{36}$Mg 
are plotted in the upper panel, while unperturbed two-quasiparticle 
strengths are shown in the lower panel. 
The arrows beside the abscissa axis indicate the neutron threshold energy 
$E_{\mathrm{th}}=4.06$ MeV (1qp continuum) and 4.12 MeV (2qp continuum).  
{\it Right}: 
Two-quasiparticle excitations generating the lowest $K^{\pi}=0^{+}$ mode 
at 3.4 MeV in $^{36}$Mg.  
}
  \end{center}
\label{36Mg_0+}
\end{figure}

\begin{table}[tp]
\caption{QRPA amplitudes of the $K^{\pi}=0^{+}$ mode at 3.4 MeV in $^{36}$Mg. 
This mode has $B(E2)=8.1 ~e^{2}$fm$^{4}$, 
$B(Q^{\nu}$2)=104 fm$^{4}$, $B(Q^{\mathrm{IS}}$2)=170 fm$^{4}$, 
and $\sum|g_{\alpha\beta}|^{2}=3.91 \times 10^{-2}$.  
Only components with $|f_{\alpha\beta}|^{2}-|g_{\alpha\beta}|^{2} > 0.01$ 
are listed. 
}
\label{36Mg_amplitude0+}
\begin{center} 
\begin{tabular}{c|c|c|c|c|c|c}
 & $\alpha$ & $\beta$ & $E_{\alpha}+E_{\beta}$ & $|f_{\alpha \beta}|^{2}-|g_{\alpha\beta}|^{2}$ & 
$Q_{20,\alpha\beta}^{(\mathrm{uv})}$ & $M_{20,\alpha\beta}^{(\mathrm{uv})}$ \\
 &  &  & (MeV) &  & (fm$^{2}$) & (fm$^{2}$)  \\ \hline \hline
(a) & $\nu$[310]1/2 & $\nu$[310]1/2 & 4.06 & 0.071 & 5.80 & $-1.58$ \\
(b) & $\nu$[321]3/2 & $\nu$[321]3/2 & 4.48 & 0.098 & 4.60 & $-1.61$ \\
(c) & $\nu$[312]5/2 & $\nu$[312]5/2 & 4.87 & 0.227 & 0.714 & 0.347 \\
(d) & $\nu$[310]1/2 & $\nu$[330]1/2 & 4.91 & 0.211 & $-3.08$ & $-2.11$ \\
(e) & $\nu$[301]1/2 & $\nu$[310]1/2 & 5.69 & 0.033 & 2.02 & $-0.511$ \\
(f) & $\nu$[330]1/2 & $\nu$[330]1/2 & 5.76 & 0.116 & 3.98 & $-1.50$  \\
(g) & $\nu$[202]3/2 & $\nu$[202]3/2 & 5.79 & 0.046 & $-1.47$ & $-0.271$ \\
(h) & $\nu$[303]7/2 & $\nu$[303]7/2 & 7.67 & 0.049 & $-1.82$ & $-0.411$  \\ 
\hline
(i) & $\pi$[211]1/2 & $\pi$[220]1/2 & 6.44 & 0.054 & $-0.251$ & -0.599 \\ 
\end{tabular}
\end{center} 
\end{table}

The quadrupole transition strengths calculated for $^{36}$Mg are 
displayed in Fig.~6. %\ref{36Mg_0+}. 
We notice a prominent peak at about 3.4 MeV 
below the one-neutron threshold energy (4.1 MeV), 
which possesses a strongly enhanced transition strength of about 24 W.u. 
(1 W.u. $\simeq$ 7.1 fm$^{4}$ for $^{36}$Mg). 
This peak exhibits a clear character of collective vibration: 
As seen from Table~\ref{36Mg_amplitude0+}, 
this collective mode is created by coherent neutron excitations. 
Its main components are the particle-hole type  
$\nu[310]1/2 \otimes \nu[330]1/2$ and 
$\nu[301]1/2 \otimes \nu[310]1/2$ excitations 
and the particle-particle type 
$\nu[312]5/2 \otimes \nu[312]5/2$ and 
$\nu[321]3/2 \otimes \nu[321]3/2$ excitations.
These particle-particle type and particle-hole type excitations are 
coherently superposed to generate this collective neutron mode. 

\subsection{$K^{\pi}=2^{+}$ modes}

\begin{figure}[tp]
  \begin{center}
    \begin{tabular}{cc}
	\includegraphics[height=8cm]{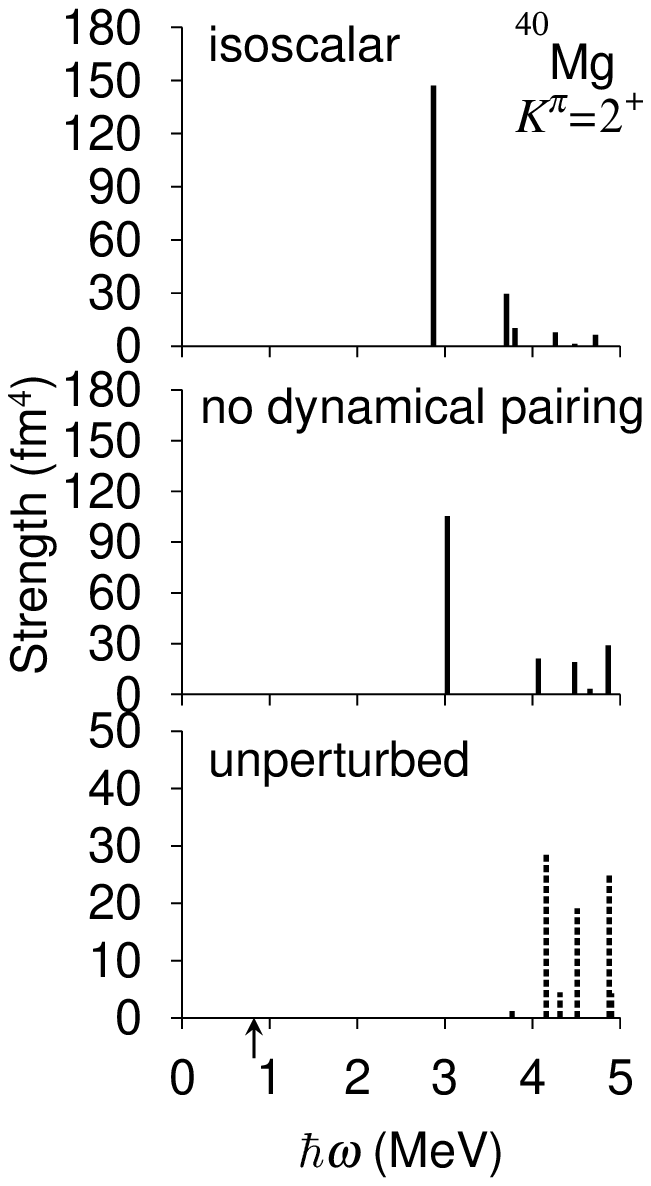}
	\includegraphics[height=8cm]{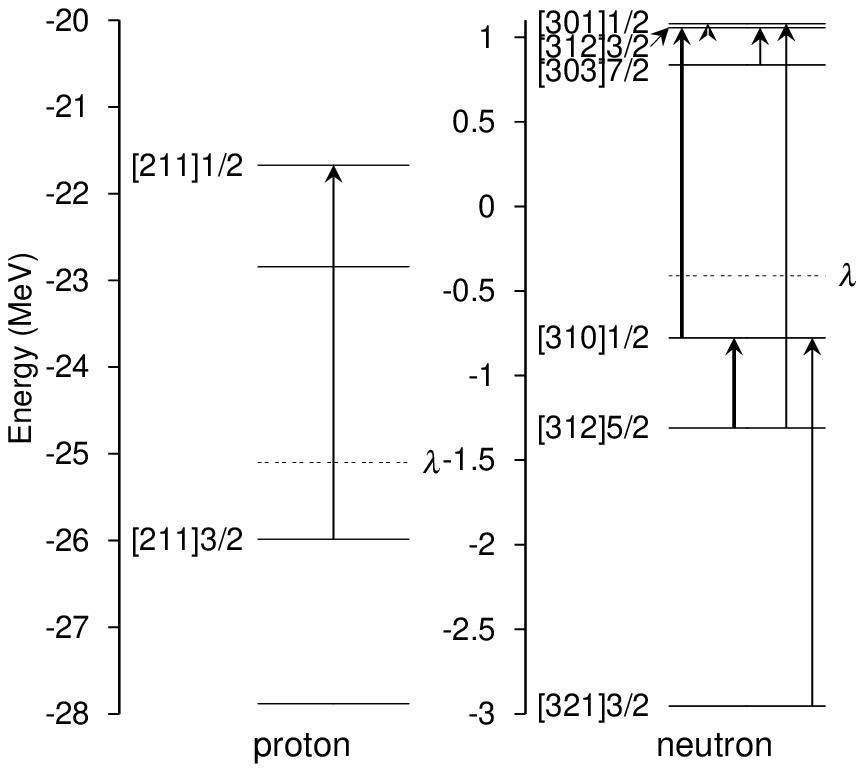}
    \end{tabular}
\caption{{\it Left}: 
Isoscalar quadrupole transition strengths $B(Q^{\mathrm{IS}}2)$ 
for the $K^{\pi}=2^{+}$ excitations in ${}^{40}$Mg. 
Results of the QRPA calculation with and without including 
the dynamical pairing effects are plotted in the upper and middle panels, 
respectively, while unperturbed two-quasiparticle strengths 
are shown in the lower panel. Notice that different scale is used for 
the unperturbed strengths.
The arrow beside the abscissa axis indicates 
the neutron threshold energy $2|\lambda|=0.82$ MeV. 
{\it Right}: 
Two-quasiparticle excitations generating the lowest $K^{\pi}=2^{+}$ mode 
at 2.9 MeV. 
Two-quasiparticle excitations 
satisfying the asymptotic selection rule for the $\gamma$ vibration 
($\Delta N=0, \Delta n_{3}=0, \Delta \Lambda=2$) are drawn by thick arrows. 
}
\label{40Mg_2+}
\end{center}
\end{figure}

\begin{table}[tbp]
\caption{
QRPA amplitudes of the $K^{\pi}=2^{+}$ mode at 2.9 MeV in $^{40}$Mg. 
This mode has $B(E2)=11.7~e^{2}$fm$^{4}$, 
$B(Q^{\nu}$2)=75.7 fm$^{4}$, $B(Q^{\mathrm{IS}}$2)=147 fm$^{4}$, 
and $\sum|g_{\alpha\beta}|^{2}=6.73 \times 10^{-2}$. 
Only components with $|f_{\alpha \beta}|^{2}-|g_{\alpha\beta}|^{2} > 0.01$ 
are listed. The label $\nu 1/2^{-}$ denotes a discretized non-resonant 
continuum state.
}
\label{40Mg_amplitude2+}
\begin{center} 
\begin{tabular}{c|c|c|c|c|c|c}
 & $\alpha$ & $\beta$ & $E_{\alpha}+E_{\beta}$ & $|f_{\alpha \beta}|^{2}-|g_{\alpha\beta}|^{2}$ & $Q_{22,\alpha\beta}^{(\mathrm{uv})}$ & $M_{22,\alpha\beta}^{(\mathrm{uv})}$  \\
 &  &  & (MeV) &  & (fm$^{2}$) & (fm$^{2}$)  \\ \hline \hline
(a) & $\nu$[312]3/2 & $\nu$[310]1/2 & 3.77 & 0.013 & 1.22    & $-0.145$  \\ 
(b) & $\nu$[301]1/2 & $\nu$[312]3/2 & 4.16 & 0.098 & $-5.37$ & $-1.75$   \\
(c) & $\nu$[310]1/2 & $\nu$[312]5/2 & 4.51 & 0.085 & $-4.37$ & $-1.34$   \\
(d) & $\nu$[312]3/2 & $\nu$[303]7/2 & 4.88 & 0.011 & $-5.03$ & $-0.454$  \\
(e) & $\nu$[301]1/2 & $\nu$[312]5/2 & 4.90 & 0.016 & $-2.07$ & $-0.296$  \\
(f) & $\nu$[310]1/2 & $\nu$[321]3/2 & 5.34 & 0.047 & $-2.67$ & $-0.663$  \\ 
(g) & $\nu 1/2^{-}$ & $\nu$[312]5/2 & 6.96 & 0.015 & 1.93    & $-0.298$  \\
(h) & $\nu 1/2^{-}$ & $\nu$[321]3/2 & 7.28 & 0.018 & 1.46    & $-0.265$  \\ 
\hline
(i) & $\pi$[211]1/2 & $\pi$[211]3/2 & 4.32 & 0.596 & $-2.11$ & $-2.02$   \\ 
\end{tabular}
\end{center} 
\end{table}

Let us now turn to the $K^{\pi}=2^{+}$ excitation modes. 
The quadrupole transition strengths calculated for $^{40}$Mg 
are displayed in Fig.~\ref{40Mg_2+}.
We notice a prominent peak at about 2.8 MeV which possesses 
strongly enhanced transition strength of about 19 W.u. 
The QRPA amplitudes of this excitation are listed 
in Table \ref{40Mg_amplitude2+}. From this Table, we see that
this peak represents a collective excitation consisting of 
a coherent superposition of the proton particle-hole excitation 
from the $[211]3/2$ level to the $[211]1/2$ level 
and a number of neutron two-quasiparticle excitations. 
Similarly to the $K^{\pi}=0^{+}$ excitation modes discussed in the previous 
subsection, the asymptotic selection rule 
($\Delta N=0, \Delta n_{3}=0, \Delta \Lambda=2$)
well known for the $\gamma$ vibrations is violated for the neutron excitations,
because these levels are loosely bound or resonances and their 
quasiparticle wave functions are significantly extended outside of the nucleus.
On the other hand, proton particle-hole excitations satisfy the selection rule 
because they are deeply bound.
We also show in Fig.~\ref{40Mg_2+} the result of QRPA calculation
where the residual pairing interaction is turned off. 
Comparing with the full QRPA result, 
we see that the transition strength is reduced about 30\%. 
Thus, the dynamical pairing effect is important, though its effect is 
weaker than for the $K^{\pi}=0^{+}$ mode. This is because 
the $K^{\pi}=2^{+}$ mode consists of both proton and neutron excitations and
the pairing is effective only for neutrons.

\begin{figure}[tp]
  \begin{center}
    \begin{tabular}{cc}
	\includegraphics[height=8cm]{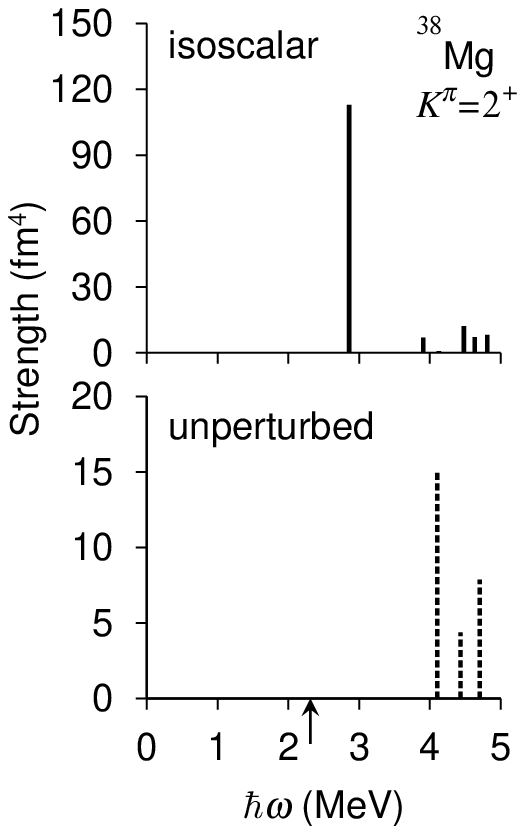}
	\includegraphics[height=8cm]{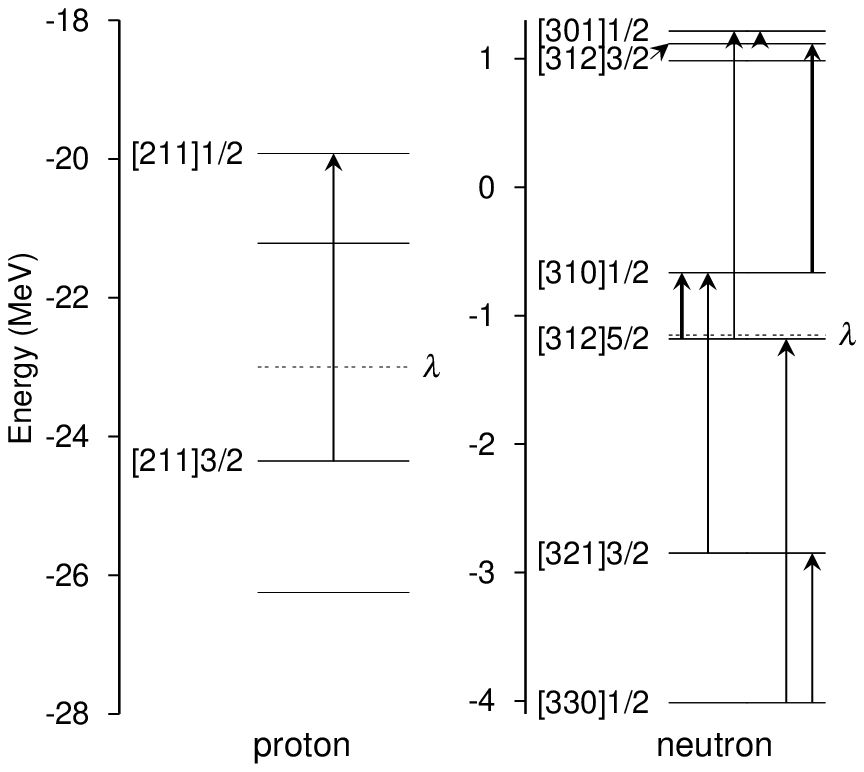}
    \end{tabular}
    \caption{{\it Left}: 
Isoscalar quadrupole transition strengths 
$B(Q^{\mathrm{IS}}2)$ for the $K^{\pi}=2^{+}$ excitations in ${}^{38}$Mg 
are plotted in the upper panel, while unperturbed two-quasiparticle 
strengths are shown in the lower panel.  
The arrow beside the abscissa axis indicates 
the neutron threshold energy $2|\lambda|=2.31$ MeV. 
{\it Right}: 
Two-quasiparticle excitations generating the lowest $K^{\pi}=2^{+}$ mode 
at 2.9 MeV in $^{38}$Mg.
}
\label{38Mg_2+}
  \end{center}
\end{figure}

\begin{table}[tp]
\caption{QRPA amplitudes of the $K^{\pi}=2^{+}$ mode at 2.9 MeV in $^{38}$Mg. 
This mode has $B(E2)=11.2 ~e^{2}$fm$^{4}$, 
$B(Q^{\nu}$2)=53.0 fm$^{4}$, $B(Q^{\mathrm{IS}}$2)=113 fm$^{4}$, 
and $\sum|g_{\alpha\beta}|^{2}=6.95 \times 10^{-2}$. 
Only components with $|f_{\alpha\beta}|^{2}-|g_{\alpha\beta}|^{2} > 0.01$ 
are listed. 
}
\label{38Mg_amplitude2+}
\begin{center} 
\begin{tabular}{c|c|c|c|c|c|c}
 & $\alpha$ & $\beta$ & $E_{\alpha}+E_{\beta}$ & 
 $|f_{\alpha \beta}|^{2}-|g_{\alpha\beta}|^{2}$ & 
$Q_{22,\alpha\beta}^{(\mathrm{uv})}$ & $M_{22,\alpha\beta}^{(\mathrm{uv})}$ \\
 &  &  & (MeV) &  & (fm$^{2}$) & (fm$^{2}$)  \\ \hline \hline
(a) & $\nu$[310]1/2 & $\nu$[312]5/2 & 4.10 & 0.146 & -3.89 & -1.54 \\
(b) & $\nu$[312]3/2 & $\nu$[310]1/2 & 4.13 & 0.016 & -0.221 & -0.032  \\
(c) & $\nu$[310]1/2 & $\nu$[321]3/2 & 4.70 & 0.108 & 2.81 & -1.05  \\
(d) & $\nu$[301]1/2 & $\nu$[312]5/2 & 5.15 & 0.010 & 1.81 & -0.204  \\
(e) & $\nu$[301]1/2 & $\nu$[312]3/2 & 5.18 & 0.031 & -3.58 & -0.690 \\
(f) & $\nu$[312]5/2 & $\nu$[330]1/2 & 6.08 & 0.017 & -1.42 & -0.239  \\
(g) & $\nu$[321]3/2 & $\nu$[330]1/2 & 6.68 & 0.029 & 1.61 & -0.344  \\ \hline
(h) & $\pi$[211]1/2 & $\pi$[211]3/2 & 4.43 & 0.541 & -2.09 & -1.94  \\ 
\end{tabular}
\end{center} 
\end{table}

\begin{figure}[tp]
  \begin{center}
    \begin{tabular}{cc}
	\includegraphics[height=8cm]{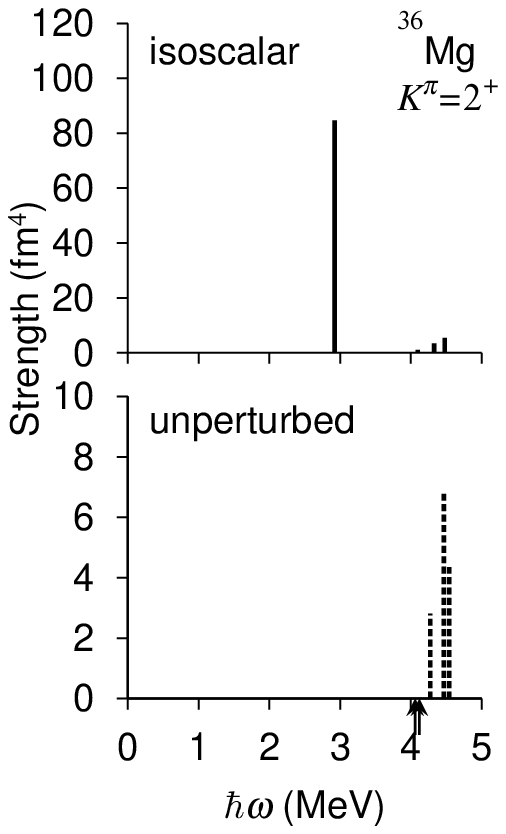}
	\includegraphics[height=8cm]{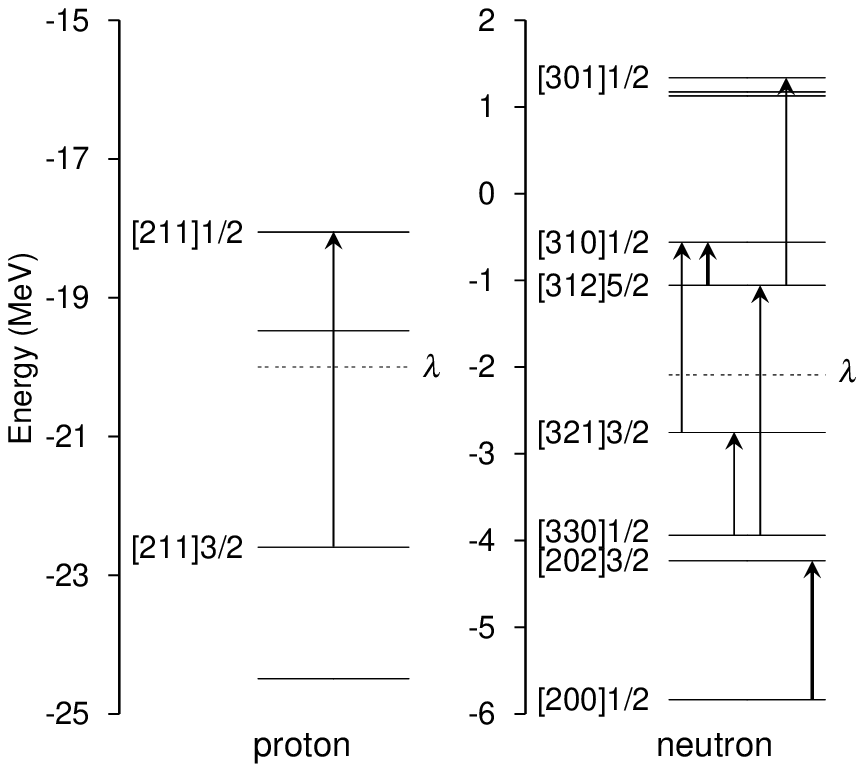}
    \end{tabular}
\caption{
{\it Left}: Isoscalar quadrupole transition strengths $B(Q^{\mathrm{IS}}2)$ 
for the $K^{\pi}=2^{+}$ excitations in $^{36}$Mg 
are plotted in the upper panel, 
while unperturbed two-quasiparticle strengths are shown in the lower panel. 
The arrows beside the abscissa axis indicate 
the threshold energies 4.06 MeV (1qp continuum) 
and 4.12 MeV (2qp continuum). 
{\it Right}: 
Two-quasiparticle excitations generating the lowest $K^{\pi}=2^{+}$ mode 
at 2.9 MeV. 
}
\label{36Mg_2+}
  \end{center}
\end{figure}

The quadrupole transition strengths calculated for $^{38}$Mg and $^{36}$Mg 
are displayed in Figs.~\ref{38Mg_2+} and \ref{36Mg_2+}, respectively.
For each case, we see a prominent peak at about 2.9 MeV 
which possesses strongly enhanced transition strength 
(about 15 W.u. and 12 W.u. for $^{38}$Mg and $^{36}$Mg, respectively).
The QRPA amplitudes of these modes are listed in Tables
\ref{38Mg_amplitude2+} and \ref{36Mg_amplitude2+}.
These modes possess essentially the same microscopic structure as 
the collective $K^{\pi}=2^{+}$ mode in $^{40}$Mg discussed above.
They also correspond to the $\gamma$ vibrational mode 
obtained in the previous QRPA calculation\cite{hag04a} for $^{38}$Mg. 
In our calculation, however, the collectivity of these modes remains almost 
the same even if we use different deformations for protons and neutrons,  
differently from Ref.~\cite{hag04a}.

\begin{table}[tp]
\caption{QRPA amplitudes of the $K^{\pi}=2^{+}$ mode at 2.9 MeV in $^{36}$Mg. 
This mode has $B(E2)=10.6 ~e^{2}$fm$^{4}$, 
$B(Q^{\nu}$2)=35.3 fm$^{4}$, $B(Q^{\mathrm{IS}}$2)=84.6 fm$^{4}$, 
and $\sum|g_{\alpha\beta}|^{2}=7.06 \times 10^{-2}$.
Only components with $|f_{\alpha\beta}|^{2}-|g_{\alpha\beta}|^{2} > 0.01$ 
are listed. 
The columns (f) and (f$^{\prime}$) are assigned the same configuration,
because we obtain two discretized continuum states associated with
the $\nu[200]1/2$ level for which $E>|\lambda|$. 
}
\label{36Mg_amplitude2+}
\begin{center} 
\begin{tabular}{c|c|c|c|c|c|c}
 & $\alpha$ & $\beta$ & $E_{\alpha}+E_{\beta}$ & $|f_{\alpha \beta}|^{2}-|g_{\alpha\beta}|^{2}$ & 
$Q_{22,\alpha\beta}^{(\mathrm{uv})}$ & $M_{22,\alpha\beta}^{(\mathrm{uv})}$ \\
 &  &  & (MeV) &  & (fm$^{2}$) & (fm$^{2}$)  \\ \hline \hline
(a) & $\nu$[310]1/2 & $\nu$[321]3/2 & 4.27 & 0.087 & 1.68 & 0.552  \\ 
(b) & $\nu$[310]1/2 & $\nu$[312]5/2 & 4.46 & 0.045 & $-2.62$ & 0.551  \\ 
(c) & $\nu$[321]3/2 & $\nu$[330]1/2 & 5.12 & 0.165 & 2.51 & 1.20  \\ 
(d) & $\nu$[312]5/2 & $\nu$[330]1/2 & 5.31 & 0.034 & $-1.36$ & 0.305  \\
(e) & $\nu$[301]1/2 & $\nu$[312]5/2 & 6.09 & 0.011 & 1.66 & 0.202  \\
(f) & $\nu$[202]3/2 & $\nu$[200]1/2 & 6.90 & 0.010 & $-2.633$ & 0.270  \\
(f$^{\prime}$) & $\nu$[202]3/2 & $\nu$[200]1/2 & 7.23 & 0.013 & 0.981  & 0.136 
\\ \hline
(g) & $\pi$[211]1/2 & $\pi$[211]3/2 & 4.54 & 0.547 & $-2.09$ & 1.87 \\ 
\end{tabular}
\end{center} 
\end{table}

\begin{figure}[tp]
\begin{center}
\includegraphics[scale=0.98]{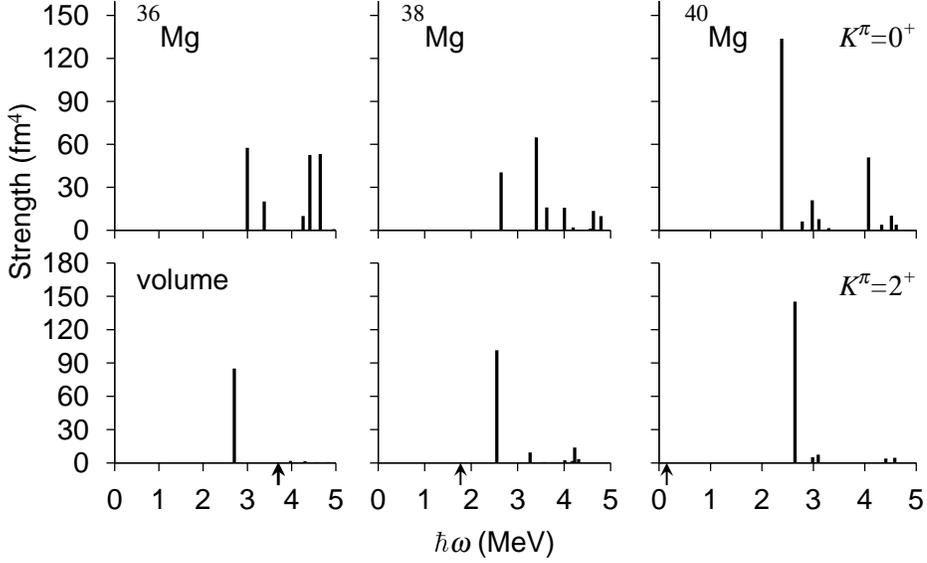}
\caption{Isoscalar quadrupole transition strengths $B(Q^{\mathrm{IS}}$2) 
for the $K=0^{+}$ excitations (upper panel) 
and the $K=2^{+}$ excitations 
(lower panel) 
built on the prolately deformed ground states of ${}^{36,38,40}$Mg. 
The QRPA calculations are made in the same way as 
in Fig.~\ref{Mg_strength}, 
except that the volume-type pairing interaction is used here. 
The arrows indicate the neutron threshold energies; 
3.69 MeV (1qp continuum) and 3.71 MeV (2qp continuum) for $^{36}$Mg, 
1.77 MeV (2qp continuum) for $^{38}$Mg, and 
0.15 MeV (2qp continuum) for ${}^{40}$Mg. 
}
\label{Mg_strength_volume}
\end{center}
\end{figure}

\subsection{Dependence on pairing interaction}

In this subsection, we examine sensitivity of the low-frequency 
$K^{\pi}=0^{+}$ and $2^{+}$ modes on 
the density dependence of the pairing interaction. 
For this purpose, we repeated the HFB and QRPA calculations using 
pairing interactions with density dependence different 
from the surface type ($\eta=1.0$ in Eq.~(\ref{eq:res_pp})); 
i.e., the mixed type ($\eta=0.5$) and the volume type ($\eta=0.0$).
Since the result for the mixed-type pairing is intermediate between those 
for the surface-type and the volume-type, we show 
in Fig.~\ref{Mg_strength_volume} only
the quadrupole transition strengths obtained using the 
volume-type pairing interaction. 
In this calculation, the pairing interaction strength  
$V_{0}=-215.0$ MeV$\cdot$fm$^{3}$ is chosen to yield approximately the same 
average pairing gaps as those for the surface type. 
Comparing with the results obtained using the surface-type pairing, 
shown in Fig.~\ref{40Mg_0+}, 
we see that the transition strengths for the $K^{\pi}=0^{+}$ collective modes 
are appreciably reduced, while those for the $K^{\pi}=2^{+}$ collective modes
are almost the same. We have checked that, although the strengths are reduced, 
the microscopic structure of these collective modes are basically the same as 
discussed above on the basis of the results of calculation using the 
surface-type pairing interaction. 
Thus, we can say that the quadrupole transition strengths for the low-frequency 
$K^{\pi}=0^{+}$ collective modes are especially sensitive to the density 
dependence of the pairing interaction. 
Such a sensitivity has been stressed also by Matsuo et al. 
in their continuum QRPA calculations for $E1$ strength functions 
in neutron rich O, Ca and Ni isotopes\cite{mat05}.

\section{Concluding remarks}

We have carried out the QRPA calculations on the basis of the deformed 
WS plus HFB mean field in the coordinate representation, and obtained 
the low-frequency $K^{\pi}=0^{+}$ and $2^{+}$ collective modes in 
deformed ${}^{36,38,40}$Mg  close to the neutron drip line.
It has been shown that these modes possess very strong isoscalar quadrupole 
transition strengths. One of the reasons of this enhancement is that the
quasiparticle wave functions participating in these collective excitations 
have spatially extended structure.
The other reason is that the residual pairing interactions, in addition to 
the particle-hole type residual interactions, enhance the collectivity of 
these modes. 
The result of the present calculation suggests that the low-frequency 
$K^{\pi}=0^{+}$ collective mode is a particularly sensitive indicator of 
the nature of pairing correlations in nuclei close to the neutron drip line. 

This paper should be regarded as an exploratory work toward understanding 
low-frequency collective modes of excitation in unstable nuclei 
close to the neutron drip line.
It is certainly desirable to improve the treatment of the continuum 
at least in the following points. 
First, one may try to use a smaller mesh size and a larger box 
by implementing an adaptive coordinate method\cite{nak05}.  
Second, one may try to take into account the width of resonance 
by employing Gamow states as basis of the QRPA calculation\cite{cur89}.
The result of the present work indicates that
calculations using such an improved framework will be 
very interesting and worthwhile. We plan to attack this subject in future.

\section*{Acknowledgment}
This work was done as a part of the Japan-U.S. Cooperative Science Program
``Mean-Field Approach to Collective Excitations in Unstable Medium-Mass 
and Heavy Nuclei" during the academic year 2003-2004, 
and we acknowledge useful discussions with the member of this project. 
One of the authors (M.Y.) is grateful for the financial assistance
from the Special Postdoctoral Researcher Program of RIKEN.
The numerical calculations were performed on the NEC SX-8 
and SX-5 supercomputers at Yukawa Institute for Theoretical Physics, 
Kyoto University and 
NEC SX-5 supercomputer at Research Center for Nuclear Physics, Osaka University.


\appendix
\section{Eigenphase sum for single-particle resonance states}

We examine properties of three single-particle states in the continuum, 
which play a key role in generating the low-lying excitations modes 
in $^{36,38,40}$Mg.
The resonance energy and width in a deformed potential can be estimated
using the eigenphase sum $\Delta(E)$. It is defined in terms
of the eigenvalues of the scattering matrix (S-matrix) as
\begin{equation}
(U^{\dagger}SU)_{aa^{\prime}}=e^{2i\delta_{a}(E)}\delta_{aa^{\prime}}, 
\hspace{0.4cm}
\Delta (E)=\sum_{a}\delta_{a}(E).
\end{equation}

We evaluate the eigenphase sum for three states 
following the procedure of Ref.~\cite{eigenphase}. 
The resonance energy and width are identified with 
the peak energy of $\frac{1}{\pi} d\Delta(E)/dE$ and its FWHM, 
respectively~\cite{gra01,mut01}. 
This evaluation is in good correspondence with another definition of 
the resonance; the Gamow state in a deformed potential~\cite{yos05b} 
which represents the pole of the $S-$matrix in the complex momentum plane.

\begin{figure}[tp]
\begin{center}
\begin{tabular}{ccc}
\includegraphics[height=6.5cm]{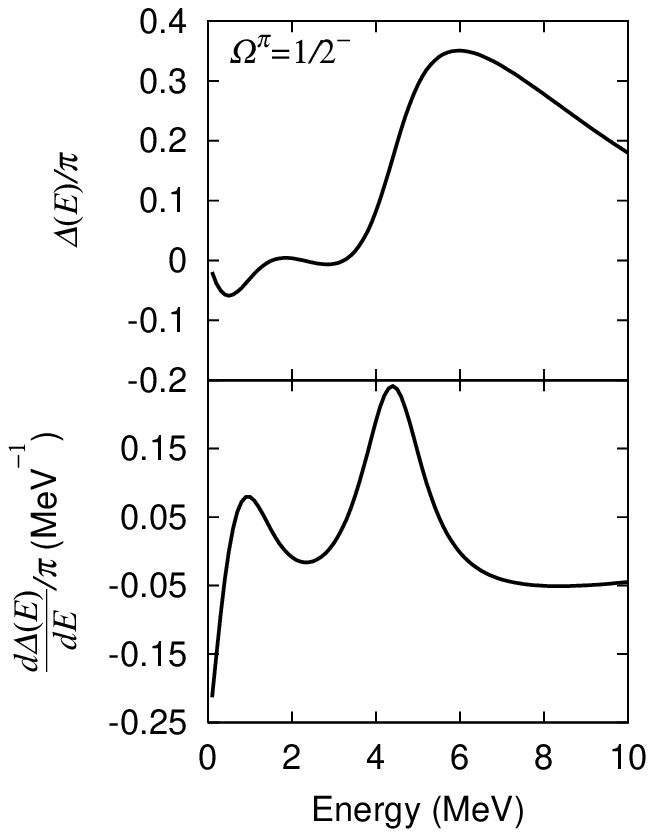}
\includegraphics[height=6.5cm]{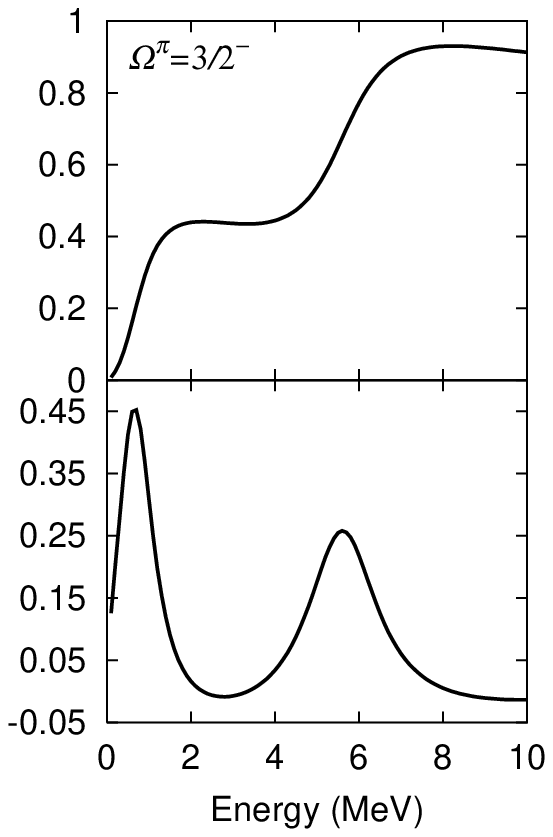}
\includegraphics[height=6.5cm]{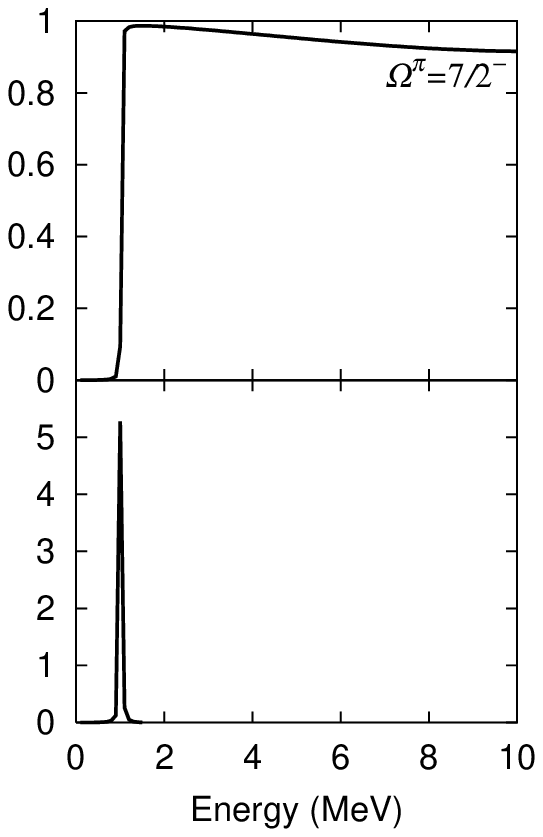}
\end{tabular}
\caption{
The eigenphase sum (upper panel) and 
its derivative (lower panel) for the 
$\Omega^{\pi}=1/2^{-}, 3/2^{-}$ and $7/2^{-}$ states
in $^{40}$Mg are plotted as functions of energy.
}
\label{40Mg_phase}
\end{center}
\end{figure}

The result of this calculation, presented in Fig.\ref{40Mg_phase}, 
indicates that the [301]1/2 and [312]3/2 states can be regarded 
as resonances with rather large widths; 
their energies are $0.53-i0.46$ (MeV) and $0.42-i0.33$ (MeV), respectively. 
On the other hand,
the [303]7/2 state is evaluated as a narrow resonance 
with energy $0.44-i0.0005$ (MeV). 
Obviously, the small width is due to its high centrifugal barrier.

\begin{thebibliography}{00}
%%%%%%%%%%%%%%%%%%%%%%%%%%%%%%%%%%%%%%%%%%%%%%%%%%%%%%%%%%%%%

\bibitem{tan01}
I.~Tanihata (Ed), 
Nucl. Phys. A 693 (2001) Nos. 1, 2.

\bibitem{hor01}
H.~Horiuchi, T.~Otsuka, Y.~Suzuki (Eds.),
Prog. Theor. Phys. Suppl. No.142 (2001).

\bibitem{hag02a}
K.~Hagino, H.~Horiuchi, M.~Matsuo, I.~Tanihata (Eds.), 
Prog. Theor. Phys. Suppl. No.146 (2002).

%%%%%%%% RPA approach %%%%%%%%%%%%%%%%%%%%%%%%%%%%%

\bibitem{ham96}
I.~Hamamoto, H.~Sagawa, X.~Z.~Zhang,  
Phys. Rev. C 53 (1996) 765;
ibid. 55 (1997) 2361;
ibid. 56 (1997) 3121;
ibid. 57 (1998) R1064;
ibid. 64 (2001) 024313.

\bibitem{ham99}
I.~Hamamoto, H.~Sagawa, Phys. Rev. C 60 (1999) 064314;
ibid. 62 (2000) 024319;
ibid. 66 (2002) 044315.

\bibitem{shl03}
S.~Shlomo, B.~Agrawal, Nucl. Phys. A 722 (2003) 98c.

%%%%%%% QRPA approach %%%%%%%%%%%%%%%%%%%%%%%%%%%%%%%
\bibitem{mat01}
M.~Matsuo, Nucl. Phys. A 696 (2001) 371.

\bibitem{hag01}
K.~Hagino, H.~Sagawa, Nucl. Phys. A 695 (2001) 82.

\bibitem{ben02}
M.~Bender, J.~Dobaczewski, J.~Engel, W.~Nazarewicz, 
Phys. Rev. C 65 (2002) 054322.

\bibitem{kha02}
E.~Khan, N.~Sandulescu, M.~Grasso, N.~Van Giai, 
Phys. Rev. C 66 (2002) 024309.

\bibitem{yam04}
M.~Yamagami, N.~Van Giai, Phys. Rev. C 69 (2004) 034301.

\bibitem{ter05}
J.~Terasaki, J.~Engel, M.~Bender, J.~Dobaczewski, W.~Nazarewicz, 
M.~Stoitsov, Phys. Rev. C 71 (2005) 034310.

\bibitem{mat05}
M.~Matsuo, K.~Mizuyama, Y.~Serizawa, Phys. Rev. C 71 (2005) 064326.

%%%%%%% Relativistic %%%%%%%%%%%%%%%%
%% RPA %%
\bibitem{vre01}
D.~Vretenar, N.~Paar, P.~Ring, G.~A.~Lalazissis, 
Nucl. Phys. A 692 (2001) 496

%% QRPA %%
\bibitem{paa03}
N.~Paar, P.~Ring, T.~Nik\v si\'c, D.~Vretenar, 
Phys. Rev. C 67 (2003) 034312.

\bibitem{paa04}
N.~Paar, T.~Nik\v si\'c, D.~Vretenar, P.~Ring, 
Phys. Rev. C 69 (2004) 054303.

\bibitem{paa05}
N.~Paar, T.~Nik\v si\'c, D.~Vretenar, P.~Ring, 
Phys. Lett. B 606 (2005) 288.

\bibitem{cao05}
L.~G.~Cao, Z.~Y.~Ma, 
Phys. Rev. C 71 (2005) 034305.

\bibitem{vre05}
D.~Vretenar, A.~V.~Afanasjev, G.~A.~Lalazissis, P.~Ring, 
Phys. Rep. 409 (2005) 101.

%%%%%% Finite-range %%%%%%%%%%%%%%%%%%%%%%%%%%%%%%%%%
\bibitem{gia03}
G.~Giambrone, S.~Scheit, F.~Barranco, P.~F.~Bortignon,
G.~Col\`o, D.~Sarchi, E.~Vigezzi, Nucl. Phys. A 726 (2003) 3.

\bibitem{per05}
S.~P\'eru, J.~F.~Berger, P.~F.~Bortignon, 
Eur. Phys. J. A 26 (2005) 25.
 
%%%%%%%%%%%%%%%%%%%%%%%%%%%%%%%%%%%%%%%%%%%%%%%%%%%%%%%
\bibitem{sar04}
D.~Sarchi, P.~F.~Bortignon, G.~Col\'o, 
Phys. Lett. B 601 (2004) 27.

%%%%%%%%%%%%%%%%%%%%%%%%%%%%%%%%%%%%%%%%%%%%%%%%%%%%%%
\bibitem{ben03}
M.~Bender, P-H.~Heenen, P-G.~Reinhard, 
Rev. Mod. Phys. 75 (2003) 121.

%%%%%%% deformed RPA %%%%%%%%%%%%%%%%%%%%%%%%%%%%%%%%%
\bibitem{nak05}
T.~Nakatsukasa, K.~Yabana, Phys. Rev. C 71 (2005) 024301.

\bibitem{ina05}
T.~Inakura, H.~Imagawa, Y.~Hashimoto, S.~Mizutori, 
M.~Yamagami, K.~Matsuyanagi, Nucl. Phys. A 768 (2006) 61

\bibitem{lem68}
R.~H.~Lemmer, M.~V$\acute{\mathrm{e}}$n$\acute{\mathrm{e}}$roni, 
Phys. Rev. 170 (1968) 883.

\bibitem{mut02}
A.~Muta, J-I.~Iwata, Y.~Hashimoto, K.~Yabana, 
Prog. Theor. Phys. 108 (2002) 1065.

\bibitem{ima03}
H.~Imagawa, Y.~Hashimoto, Phys. Rev. C 67 (2003) 037302.

\bibitem{yos05}  
K.~Yoshida, M.~Yamagami, K.~Matsuyanagi,  
Prog. Theor. Phys. 116 (2005) 1251.

\bibitem{urk01}
P.~Urkedal, X.~Z.~Zhang, I.~Hamamoto, 
Phys. Rev. C 64 (2001) 054304.

\bibitem{alv04}
R.~\'Alvarez-Rodr\'{\i}guez, P.~Sarriguren, E.~Moya de Guerra, L.~Pacearescu, A.~Faessler, F.~\v Simkovic, 
Phys. Rev. C 70 (2004) 064309 and references therein.

\bibitem{hag04a}
K.~Hagino, N.~Van Giai, H.~Sagawa, Nucl. Phys. A 731 (2004) 264.

%%%%%%%%%%%%%%%%%%%%%%%%%%%%%%%%%%%%%%%%%%%%%%%%%%%%%%%%%%
\bibitem{bul80}
A.~Bulgac, Preprint No. FT-194-1980, 
Institute of Atomic Physics, Bucharest, 1980. 
[arXiv:nucl-th/9907088]

\bibitem{dob84}
J.~Dobaczewski, H.~Flocard, J.~Treiner,
Nucl. Phys. A 422 (1984) 103.

%%%%%%%%%%%%%%%%%%%%%%%%%%%%%%%%%%%%%%%%%%%%%%%%%%%%%%%%%%%%%%%%%%%%%%
\bibitem{smo93}
R.~Smola\'nczuk, J.~Dobaczewski, Phys. Rev. C 48 (1993) R2166. 

\bibitem{dob94}
J.~Dobaczewski, I.~Hamamoto, W.~Nazarewicz, J.~A.~Sheikh, 
Phys. Rev. Lett. 72 (1994) 981.

\bibitem{dob96}
J.~Dobaczewski, W.~Nazarewicz, T.~R.~Werner, J.~F.~Berger, C.~R.~Chinn, 
J.~Decharg\'e, 
Phys. Rev. C 53 (1996) 2809.

\bibitem{ben99}
K.~Bennacuer, J.~Dobaczewski, M.~P{\l}oszajczak, 
Phys. Rev. C 60 (1999) 034308.

%%%%%%%%%%%%%%%%%%%%%%%%%%%%%%%%%%%%%%%%%%%%%%%%%%%%%%%%%%%%%%%%%%%%%
\bibitem{yam05}
M.~Yamagami, Phys. Rev. C 72 (2005) 064308.

% SHFB
\bibitem{ter97}
J.~Terasaki, H.~Flocard, P.~-H.~Heenen, P.~Bonche, 
Nucl. Phys. A 621 (1997) 706.

\bibitem{sto03}
M.~V.~Stoitsov, J.~Dobaczewski, W.~Nazarewicz, S.~Pittel, D.~J.~Dean, 
Phys. Rev. C 68 (2003) 054312.

%Gogny-HFB
\bibitem{rod02}
R.~Rodor\'\i guez-Guzm\'an, J.~L.~Egido, L.~M.~Robledo, 
Nucl. Phys. A 709 (2002) 201.

% shell model
\bibitem{cau04}
E.~Caurier, F.~Nowacki, A.~Poves, Nucl. Phys. A 742 (2004) 14.

\bibitem{rei99}
P.-G.~Reinhard, D.~J.~Dean, W.~Nazarewicz, J.~Dobaczewski, J.~A.~Maruhn, 
M.~R.~Strayer, 
Phys. Rev. C 60 (1999) 014316 and references therein.

%Nilsson 2005
\bibitem{yos05_nil}
K.~Yoshida, M.~Yamagami, K.~Matsuyanagi, 
{\it Proc. Int. Conference on Finite Fermionic Systems 
-- Nilsson Model 50 Years}, Lund, Sweden, 14-18 June, 2005, 
arXiv:nucl-th/0507047, Physica Scipta, in press.

\bibitem{obe03}
E.~Teran, V.~E.~Oberacker, A.~S.~Umar,
Phys. Rev. C 67 (2003) 064314.

\bibitem{ber91}
G.~F.~Bertsch, H.~Esbensen, Annu. Phys. 209 (1991) 327.

\bibitem{ter95}
J.~Terasaki, P.-H.~Heenen, P.~Bonche, J.~Dobaczewski, H.~Flocard, 
Nucl. Phys. A 593 (1995) 1.

%%%%%%%%%%%%%%%%%%%%%%%%%%%%%%%%%%%%%%%%%%%%%%%%%%%%%%%%%%%%%%%%%%%%%%%%%%%
\bibitem{row70}
D.~J.~Rowe, {\it Nuclear Collective Motion}, (Methuen and Co. Ltd., 1970)

\bibitem{sho75}
S.~Shlomo, G.~F.~Bertsch, Nucl. Phys. A 243 (1975) 507.

\bibitem{BM2}
A.~Bohr, B.~R.~Motteleson, 
{\it Nuclear Structure}, vol.~II (Benjamin, 1975).


%%%%%%%%%%%%%%%%%%%%%%%%%%%%%%%%%%%%%%%%%%%%%%%%%%%%%%%%%%%%%%%%%%%%%%%%%%%

\bibitem{eigenphase}
K.~Hagino, Nguyen Van Giai, 
Nucl. Phys. A 735 (2004) 55.

\bibitem{gra01}
M.~Grasso, N.~Sandulescu, Nguyen Van Giai, R.~J.~Liotta,
Phys. Rev. C 64 (2001) 064321.

%\bibitem{Hazi}
%A.~U.~Hazi, 
%Phys. Rev. A 19 (1979) 920.

\bibitem{mut01}
A.~Muta, T.~Otsuka, Prog. Theor. Phys. Suppl.
No.142 (2001) 355.

\bibitem{yos05b}
K.~Yoshida, K.~Hagino, Phys. Rev. C 72 (2005) 064311.

\bibitem{cur89}
P.~Curutchet, T.~Vertse, R.~J.~Liotta, Phys. Rev. C 39 (1989) 1020.


%%%%%%%%%%%%%%%%%%%%%%%%%%%%%%%%%%%%%%%%%%%%%%%%%%%%%%%%%%%%%%%%%%%%%%%%%



% notes:

% \bibitem{label2}
% If there is a note, it should come last:
% \bibitem{label3} \note
% \end{subbibitems}
\end{thebibliography}
\end{document}